%% file: wvz.tex
\documentclass[a4paper,11pt]{article} 
\pdfoutput=1
\usepackage{jheppub}
\usepackage{slashed}
\usepackage[dvipsnames,table]{xcolor}
\usepackage{hyperref}
\usepackage{xspace}
\usepackage[tight]{subfigure}
\usepackage{amsmath}
\usepackage{amssymb}
\usepackage{amsfonts}
\usepackage{mathrsfs}
\usepackage{comment}
\usepackage{afterpage}
\usepackage{verbatim}
\usepackage{booktabs}
\usepackage{array}
\usepackage[title,titletoc]{appendix}
\usepackage{tabularx}
\newcolumntype{C}[1]{>{\centering\arraybackslash}p{#1}}
\usepackage{scalerel}

\input{macros.tex}

\usepackage{pifont}
\usepackage[group-digits=false]{siunitx}
\usepackage{pgfplotstable}
\usepackage{booktabs}
\usepackage{array}
\usepackage{colortbl}

\title{NLO corrections to triple vector-boson production in final states with three charged leptons and two jets}
\author{Ansgar Denner,$^{a}$}
\author{Daniele Lombardi,$^{a}$}
\author{Santiago Lopez Portillo Chavez,$^{a}$ and}
\author{Giovanni Pelliccioli$^{b}$}
\affiliation{$^{a}$Institut f\"ur Theoretische Physik und Astrophysik, Universit\"at W\"urzburg, 97074 W\"urzburg, Germany}
\affiliation[]{$^{b}$Max-Planck-Institut f\"ur Physik, Boltzmannstra{\ss}e 8, 85748 Garching, Germany}
\emailAdd{ansgar.denner@uni-wuerzburg.de}
\emailAdd{daniele.lombardi@uni-wuerzburg.de}
\emailAdd{santiago.lopez-portillo-chavez@uni-wuerzburg.de}
\emailAdd{gpellicc@mpp.mpg.de}
\date{\draftdate}
\abstract{
  Tri-boson production together with vector-boson scattering is a  privileged channel to
  study the electroweak structure of the Standard Model. Upcoming LHC
  running stages will allow to measure these processes
  at unprecedented accuracy and for all possible final states, which requires to push theory predictions
  to still unexplored frontiers.
  In this work we present the first calculation for the process $\Pp\Pp\to\mu^+\mu^-\Pe^+\nu_\Pe\,\Pj\,\Pj$ at the LHC
  in a tri-boson phase space. We evaluate the three LO contributions, namely the $\mathcal{O}(\alpha^6)$, which contains the genuine tri-boson
  signature, along with the $\mathcal{O}(\as\alpha^5)$ and $\mathcal{O}(\as^2\alpha^4)$, and the two 
  $\mathcal{O}(\alpha^7)$ and $\mathcal{O}(\as\alpha^6)$ NLO corrections. The calculation is based on full Standard-Model matrix elements,
  including all resonant and non-resonant terms, complete spin correlations and interference effects.
  Integrated and differential cross sections are presented for a fiducial region inspired by High Luminosity LHC
  prospect studies.
  We find electroweak corrections of $-14\%$ for the fiducial cross
  section, almost twice as large as for other tri-boson processes.
}


\begin{document}
%%%%%%%%%%%%%%%%%%%%%%%%%%%%%%%%%
\maketitle
%%%%%%%%%%%%%%%%%%%%%
\input{introduction}
\input{details_of_calculation}

\input{results}
\input{conclusions}

\section*{Acknowledgements}
The authors are indebted to Christopher Schwan for useful discussions
and to Mathieu Pellen for some clarifications about the results of~\citere{Denner:2019tmn}.
We would like to thank Saptaparna Bhattacharya and Andrea Sciandra for fruitful discussion
on CMS and ATLAS tri-boson analyses.
This work is supported by the German Federal Ministry for
Education and Research (BMBF) under contract no.~05H21WWCAA 
and the German Research Foundation (DFG) under reference number DE 623/8-1.
The authors acknowledge support from the COMETA COST Action CA22130.

\bibliographystyle{JHEPmod}
\bibliography{wvz}

%%%%%%%%%%%%%%%%%%%%%%%%%%%%%%%%%
\end{document}

%% file: macros.tex
\def\refeq#1{\mbox{Eq.~(\ref{#1})}}
\def\refeqs#1{\mbox{Eqs.~(\ref{#1})}}
\def\refeqf#1{\mbox{(\ref{#1})}}
\def\reffi#1{\mbox{Fig.~\ref{#1}}}
\def\reffis#1{\mbox{Figs.~\ref{#1}}}
\def\refta#1{\mbox{Table~\ref{#1}}}

\def\refse#1{\mbox{Section~\ref{#1}}}
\def\refses#1{\mbox{Sections~\ref{#1}}}

\def\citere#1{\mbox{Ref.~\cite{#1}}}
\def\citeres#1{\mbox{Refs.~\cite{#1}}}

\newcommand{\newc}{\newcommand}
\newc{\beq}{\begin{equation}}
\newc{\eeq}{\end{equation}}
\newc{\bit}{\begin{itemize}}
\newc{\eit}{\end{itemize}}
\newc{\ben}{\begin{enumerate}}
\newc{\een}{\end{enumerate}}
\newc{\bce}{\begin{center}}
\newc{\ece}{\end{center}}
\newc{\bfi}{\begin{figure}}
\newc{\efi}{\end{figure}}

\newcommand{\ri}{\mathrm i}

\newcommand{\rT}{{\mathrm{T}}}

\newcommand{\rw}{{\mathrm{w}}}

\newcommand{\ie}{{i.e.}\ }

\newcommand{\GeV}{\ensuremath{\,\text{GeV}}\xspace}
\newcommand{\TeV}{\ensuremath{\,\text{TeV}}\xspace}

\newcommand{\ab}{{\ensuremath\unskip\,\text{ab}}\xspace}

\newcommand{\Pq}{\ensuremath{q}\xspace}
\newcommand{\PH}{\ensuremath{\text{H}}\xspace}
\newcommand{\Pj}{\ensuremath{\text{j}}\xspace}
\newcommand{\Pp}{\ensuremath{\text{p}}}
\newcommand{\Pe}{\ensuremath{\text{e}}\xspace}
\newcommand{\Pb}{\ensuremath{\text{b}}\xspace}
\newcommand{\Pt}{\ensuremath{\text{t}}\xspace}
\newcommand{\Pu}{\ensuremath{\text{u}}\xspace}
\newcommand{\Pd}{\ensuremath{\text{d}}\xspace}
\newcommand{\Ps}{\ensuremath{\text{s}}\xspace}
\newcommand{\Pc}{\ensuremath{\text{c}}\xspace}
\newcommand{\Pg}{\ensuremath{\text{g}}}

\newcommand{\PW}{\ensuremath{\text{W}}\xspace}
\newcommand{\PZ}{\ensuremath{\text{Z}}\xspace}
\newcommand{\Pl}{\ell}
                                    
\newcommand{\Mt}{\ensuremath{m_\Pt}\xspace}
\newcommand{\MH}{\ensuremath{M_\PH}\xspace}

\newcommand{\MW}{\ensuremath{M_\PW}\xspace}

\newcommand{\MZ}{\ensuremath{M_\PZ}\xspace}

\newcommand{\Gt}{\ensuremath{\Gamma_\Pt}\xspace}
\newcommand{\GH}{\ensuremath{\Gamma_\PH}\xspace}

\newcommand{\alphas}{\ensuremath{\alpha_\text{s}}\xspace}
\newcommand{\gs}{\ensuremath{g_\text{s}}\xspace}

\newcommand{\MSbar}{\ensuremath{\overline{{\text{MS}}\xspace}}}

\newcommand{\order}[1]{\ensuremath{\mathcal{O}{\left(#1\right)}}\xspace}

\newcommand{\recola}{{\sc Recola}\xspace}

\newcommand{\mocanlo}{{\sc MoCaNLO}\xspace}
\newcommand{\collier}{{\sc Collier}\xspace}

\newcolumntype{.}{D{.}{.}{-1}}
\newcolumntype{d}[1]{D{.}{.}{#1}}
\colorlet{tableoverheadcolor}{gray!37.5}
\colorlet{tableheadcolor}{gray!25}
\colorlet{tablerowcolor}{gray!12.5}

\def\draftdate{\relax}
\def\mda{\relax}
\def\mua{\relax}
\def\mla{\relax}
\def\draft{
\def\thtystars{******************************}
\def\sixtystars{\thtystars\thtystars}
\typeout{}
\typeout{\sixtystars**}
\typeout{* Draft mode!
         For final version remove \protect\draft\space in source file *}
\typeout{\sixtystars**}
\typeout{}
\def\draftdate{\today}
\def\mua{\marginpar[\boldmath\hfil$\uparrow$]%
                   {\boldmath$\uparrow$\hfil}\color{black}%
                    \typeout{marginpar: $\uparrow$}\ignorespaces}
\def\mda{\color{red}\marginpar[\boldmath\hfil$\downarrow$]%
                   {\boldmath$\downarrow$\hfil}%
                    \typeout{marginpar: $\downarrow$}\ignorespaces}
\def\mla{\marginpar[\boldmath\hfil$\rightarrow$]%
                   {\boldmath$\leftarrow $\hfil}%
                    \typeout{marginpar: $\leftrightarrow$}\ignorespaces}
\def\Mua{\marginpar[\boldmath\hfil$\Uparrow$]%
                   {\boldmath$\Uparrow$\hfil}\color{black}%
                    \typeout{marginpar: $\uparrow$}\ignorespaces}
\def\Mda{\color{red}\marginpar[\boldmath\hfil$\Downarrow$]%
                   {\boldmath$\Downarrow$\hfil}%
                    \typeout{marginpar: $\downarrow$}\ignorespaces}
\def\Mla{\marginpar[\boldmath\hfil\textcolor{red}{$\Rightarrow$}]%
                   {\boldmath\textcolor{red}{$\Leftarrow $}\hfil}%
                    \typeout{marginpar: $\leftrightarrow$}\ignorespaces}
\overfullrule 5pt
\oddsidemargin 15mm
\marginparwidth 29mm
}

\let\nnb\notag

\let\Mz\MZ

\let\Mwo\MWOS
\let\Gwo\GWOS
\let\Mzo\MZOS
\let\Gzo\GZOS

\newcommand{\Mvo}{M^{\rm OS}_{V}}
\newcommand{\Mv} {M_{V}}
\newcommand{\Gvo}{\Gamma^{\rm OS}_{V}}
\newcommand{\Gv} {\Gamma_{V}}
\let\as\alphas
\newcommand{\pt}[1]{p_{\rT,{#1}}}
\newcommand{\tm}[1]{M_{\rT,{#1}}}

%% file: introduction.tex
\section{Introduction}\label{sec:intro}

The successful Standard Model (SM) of particle physics is being scrutinised
by the ongoing comparison of experimental measurements from the Large Hadron Collider (LHC) and
theoretical predictions.
Any statistically significant discrepancy between
theory and experiments can point to a still poor understanding of nature and open
the path to new physics discoveries. By now, the huge amount of data that has been collected
at the LHC continually confirmed any SM-based prediction, at least within the level of
accuracy that both the experimental and the theory community can provide so far. However,
large improvements are expected from the upcoming full Run-3 dataset and even more from
the future high-luminosity (HL) stage of the LHC, where statistical uncertainties in the
experimental measurements will be considerably reduced, and which might
shed light on still unexplored or poorly known sectors of the SM. Among these, some parts of the
electroweak (EW) sector, which is a fundamental building block of the SM, are still surprisingly
loosely constrained owing to the complexity of the measurements of processes which
are sensitive to it. Among these rare processes, vector-boson
scattering (VBS) and triple vector-boson production are the
most prominent ones to directly access triple and quartic gauge couplings
and to study the mechanism of EW symmetry breaking. Since all of these production
mechanisms have already been investigated by experimental collaborations at the LHC and will be known with a much higher precision
in the next years, a solid control on the theory predictions requires to compute these processes
at the best-possible accuracy in all of their final states.

Despite the intrinsic complexity, VBS measurements have already received remarkable
attention at the LHC in the last years especially in the fully-leptonic final state, together
with some more recent searches in the semi-leptonic final state (see for instance
\citeres{BuarqueFranzosi:2021wrv,Covarelli:2021gyz}
for a general overview on the status of VBS studies). On the other hand,
the production of three gauge bosons is a much less explored signal,
since its very low
cross section and the overwhelming background render its measurement highly elusive.
If one excludes tri-boson measurements involving at least one
photon~\cite{ATLAS:2017lpx,ATLAS:2015ify,CMS:2017tzy,ATLAS:2016qjc,CMS:2014cdf,ATLAS:2017bon},
only a few searches for triple massive-gauge-boson production have been performed so far,
specifically for $\PW^{\pm }\PW^{\pm }\PW^{\mp }$ at $8\,$TeV~\cite{ATLAS:2016jeu}
and $13\,$TeV~\cite{CMS:2019mpq} by ATLAS and CMS, respectively,
followed by evidence for massive tri-boson production in
\citere{ATLAS:2019dny} and finally its observation by the two collaborations
in~\citeres{ATLAS:2022xnu,CMS:2020hjs},
where all possible massive-boson combinations have been considered. Owing to the undoubted relevance
of this process for complementing our knowledge of the SM, new measurements are expected to
be carried out to constrain the tri-boson signal further, as confirmed by an ATLAS prospect
study for the HL phase of the LHC in~\citere{ATLAS:2018iou}.

This scenario poses new challenges for the theory community. Indeed, keeping up with the
progress achieved on the experimental side requires to improve the accuracy at which
VBS and tri-boson processes are known but also to test existing ones for 
fiducial phase-space regions which have not been considered yet. This is the case for the tri-boson
phase space, which has by now received very little attention owing to its hitherto marginal role
in experimental searches. Computations for tri-boson production mostly exist as part of full VBS ones,
to which they contribute as a background. For the latter processes many results are available
in the literature, with
state-of-the-art calculations for the fully leptonic final states
including next-to-leading order (NLO) QCD and NLO EW
corrections~\cite{Biedermann:2016yds,Denner:2019tmn,Denner:2020zit,Denner:2022pwc},
or even the complete NLO corrections to $\PW^+\PW^+$~\cite{Biedermann:2017bss} and
$\PZ\PZ$~\cite{Denner:2021hsa} scattering. 
%Results matched to parton showers have also been
%obtained for the NLO QCD corrections to VBS in different frameworks like
%MadGraph5\_aMC@NLO~\cite{Stelzer:1994ta,Alwall:2014hca}, Sherpa~\cite{Sherpa:2019gpd},
%and POWHEG-BOX~\cite{Nason:2004rx,Frixione:2007vw,Alioli:2010xd}. 
Additionally, NLO EW corrections
have been matched to a QED shower and interfaced to a QCD shower for $\PW^+\PW^+$ scattering 
in~\citere{Chiesa:2019ulk}. Even if to a lesser extent, some VBS calculations in the semi-leptonic
final state are also becoming available, by now only at leading order (LO)~\cite{Ballestrero:2008gf,Denner:2024xul}.
This rich and potentially increasing list of results for VBS should be compared to the much lower
number of studies focusing on triple massive-gauge-boson production.
In fact, while tri-boson calculations exist at NLO QCD \cite{Lazopoulos:2007ix,Binoth:2008kt}
and NLO QCD+EW \cite{Nhung:2013jta,Dittmaier:2017bnh} accuracy for
stable vector bosons or with LO decays in the
narrow-width approximation \cite{Shen:2015cwj,Shen:2016ape} since many
years, the off-shell description of such processes
has been limited to NLO QCD accuracy in the fully leptonic channel for a long time \cite{Hankele:2007sb,Campanario:2008yg},
and NLO EW corrections were only achieved a few years ago \cite{Schonherr:2018jva,Dittmaier:2019twg}.
The case of tri-boson production with one vector boson decaying hadronically has been addressed only very
recently in \citere{Denner:2024ufg}, where the full set of NLO corrections for tri-boson and $\PW\PH$
production in the $\PW^+\PW^+\Pj\Pj$ channel have been considered. In the same work
the NLO QCD contribution has been matched to the Sherpa parton shower~\cite{Sherpa:2019gpd}
for the QCD and EW production modes, where for the latter EW corrections have been included
via the EW$_{\textrm{virt}}$ approximation~\cite{Kallweit:2015dum,Gutschow:2018tuk}.
The importance of tri-boson processes as probes of possible new-physics effects beyond the SM
is witnessed by very recent LO \cite{Bellan:2023efn} and NLO QCD \cite{Celada:2024cxw} studies
in the framework of the SM effective field theory.

The production mechanism of three massive vector bosons in a non-fully-leptonic final state has a clear
overlap with VBS from a theoretical point of view. Indeed, fully off-shell theory calculations
are only sensitive to the final state of the process, while they account for all possible intermediate
resonances. The two production modes are actually defined and distinguished by the fiducial phase-space
that is used to enhance a specific signal over a certain background. As shown in~\citere{Denner:2024ufg}
in the context of tri-boson and VBS production, a change in the definition
of the fiducial volume can lead to
phenomenologically interesting outcomes, for instance a different behaviour of the EW corrections.
That already represents  a strong enough theoretical motivation to study tri-boson production for this yet little
explored semi-leptonic final state. Moreover, the latter, which has a larger cross section compared to its
fully-leptonic counterpart, is expected to become accessible at the LHC and by that time it will definitely
play a relevant role in increasing our understanding of the process.

In this work we provide one more calculation for tri-boson production with one hadronically decaying
vector boson, specifically in the $\PW^+\PZ\Pj\Pj$ channel. This
channel has been studied before in the VBS phase space in
\citere{Denner:2019tmn}. In the present work, all LO contributions to the process
have been computed, namely $\mathcal{O}(\alpha^6)$, $\mathcal{O}(\as\alpha^5)$, and
$\mathcal{O}(\as^2\alpha^4)$. Moreover, the $\mathcal{O}(\alpha^7)$ and $\mathcal{O}(\as\alpha^6)$
corrections have also been obtained exactly. A detailed description of the calculation for the different
perturbative contributions is presented in~\refse{sec:calcdetails}. Then we report numerical results
in \refse{sec:numresults}, obtained in a
phase space devised to enhance the tri-boson signal and inspired
by the HL LHC prospect studies from ATLAS
\cite{ATLAS:2018iou}. After fixing our setup in \refse{sec:input}, both inclusive and
differential cross sections are reported in \refses{sec:integrated}
and \ref{sec:differential}, respectively. We conclude by commenting on the main outcomes of
our study in \refse{sec:conclusions}.

%%% Local Variables: 
%%% mode: latex
%%% TeX-master: "wvz_paper"
%%% End: 

%% file: details_of_calculation.tex
\section{Details of the calculation}\label{sec:calcdetails}

In this paper we investigate the process
\beq\label{eq:procdef}
\Pp\Pp\to\mu^+\mu^-\Pe^+\nu_\Pe\,\Pj\,\Pj\,
\eeq
at the LHC in a phase space that is devised to enhance the tri-boson production mechanism.

\begin{figure*}
  \centering
  \includegraphics[scale=0.55]{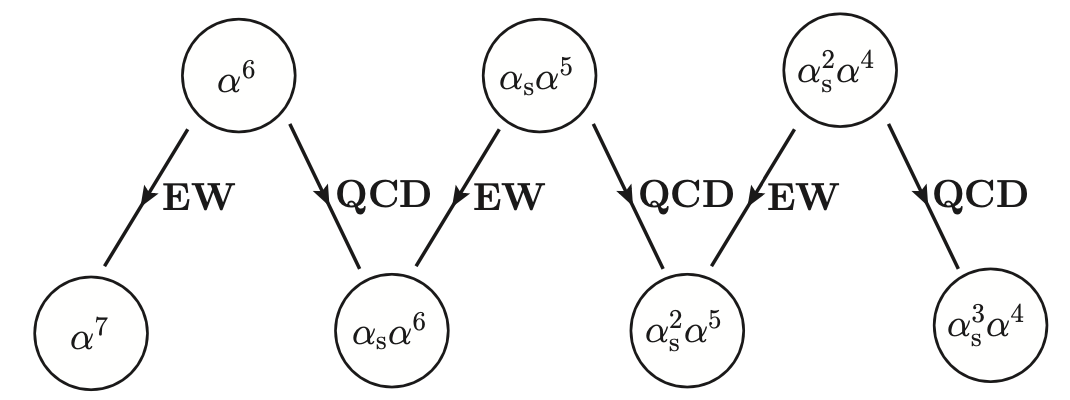}
  \caption{
    Perturbative contributions to the process $\Pp\Pp\to\mu^+\mu^-\Pe^+\nu_\Pe\,\Pj\,\Pj$.
  }\label{fig:orders}
\end{figure*}

We evaluate the process for the full set of LO contributions, namely $\mathcal{O}(\alpha^6)$,
$\mathcal{O}(\as\alpha^5)$, and $\mathcal{O}(\as^2\alpha^4)$. Only the $\mathcal{O}(\alpha^6)$
includes the tri-boson signal, whereas the remaining two orders
exclusively describe its irreducible background. We additionally
compute the two NLO corrections to the $\mathcal{O}(\alpha^6)$, which contribute at 
$\mathcal{O}(\alpha^7)$ and $\mathcal{O}(\as\alpha^6)$. For the latter order, EW corrections to $\mathcal{O}(\as\alpha^5)$
are also properly included, since they can not be disentangled from the QCD corrections to $\mathcal{O}(\alpha^6)$ (see \refse{sec:nlo2}).
That is illustrated in \reffi{fig:orders}, where we show a summary of all perturbative orders which are relevant for the process
in~\refeq{eq:procdef} up to NLO. Among these, we refrain from evaluating the
$\mathcal{O}(\as^2\alpha^5)$ and $\mathcal{O}(\as^3\alpha^4)$, since they only comprise corrections
to the tri-boson background. 

The calculation has been performed with the in-house program \mocanlo,
a multichannel Monte Carlo generator that has already proven suitable for the
evaluation of processes with high-multiplicity final states and  an
intricate resonance structure, like the one presented in this article.
%\cite{Denner:2020hgg,Denner:2021hqi,Denner:2015yca,Denner:2016jyo,Denner:2017kzu,Denner:2016wet}.
It is interfaced with \recola \cite{Actis:2012qn, Actis:2016mpe}, which 
provides the tree-level SM matrix elements together with the spin-correlated and colour-correlated
ones, needed for the definition of the unintegrated subtraction counterterms.
\recola computes all the required one-loop amplitudes using the \collier
library \cite{Denner:2016kdg} to perform the reduction and numerical
evaluation of one-loop integrals
\cite{Denner:2002ii,Denner:2005nn,Denner:2010tr}.%

We work in a five-flavour scheme and assume a 
diagonal quark-mixing matrix with unit entries throughout our calculations. We explicitly consider
a final state involving two different lepton generations. Predictions with two same-sign same-flavour
leptons can be roughly recovered by multiplying our results by appropriate factors
that account for the number of identical particles in the final state. Interference effects, which
are neglected by this procedure, are expected to be small as was
verified for fully-leptonic tri-boson processes in
\citeres{Schonherr:2018jva,Dittmaier:2019twg}.

Even though the  phase space defined in~\refse{sec:numresults} is devised to increase the
tri-boson signal, our calculation consistently includes all possible resonant and non-resonant topologies
and exactly retains interference effects. Owing to our choice of
flavour scheme, the cross section for
the process in~\refeq{eq:procdef} also receives contributions from partonic processes
with bottom quarks both as initial and as final states. In the latter
case the bottom quarks can give rise to
a top-quark resonance, which contaminates our genuine tri-boson signal in a non-negligible way.
As discussed in the following, we performed a complete LO calculation for these contributions
but do not include them in our final predictions nor evaluate them at NLO. Indeed, they should
rather be considered as part of a different LHC process, and we drop them by assuming a perfect
$\Pb$-jet tagging and veto.

Since experimental analyses often separate photon-production from jet-production processes, we do not treat
final-state photons as jets. This experimentally-motivated choice has important consequences in the treatment
of NLO corrections at $\mathcal{O}(\alphas\alpha^6)$, where different subtleties arise,
as discussed in~\refse{sec:nlo2}.

\subsection{Leading-order contributions}
\label{sec:leadingorder}

Using $g$ and $\gs$ to denote the EW and strong coupling constants, respectively,
the amplitude for the process in~\refeq{eq:procdef} receives contributions at
$\mathcal{O}(g^6)$, $\mathcal{O}(\gs g^5)$, and $\mathcal{O}(\gs^2 g^4)$.
At the squared-amplitude level three
different LO contributions are present, namely $\mathcal{O}(\alpha^6)$,
$\mathcal{O}(\as\alpha^5)$, and $\mathcal{O}(\as^2\alpha^4)$.

\subsubsection{Contributions to $\mathcal{O}(\alpha^6)$}
\label{sec:lo1}

The $\mathcal{O}(\alpha^6)$ is the one containing our signal, comprising two
leptonically-decaying gauge bosons and a hadronically-decaying one, which can either be
a $\PZ$ or a $\PW$ boson. For the specific choice of leptonic final states in~\refeq{eq:procdef},
charge conservation forces the hadronically-decaying $\PW$ boson to be negatively charged.

\begin{figure}
  \centering
  \subfigure[\label{fig:a6_q1}]{\includegraphics[scale=0.30]{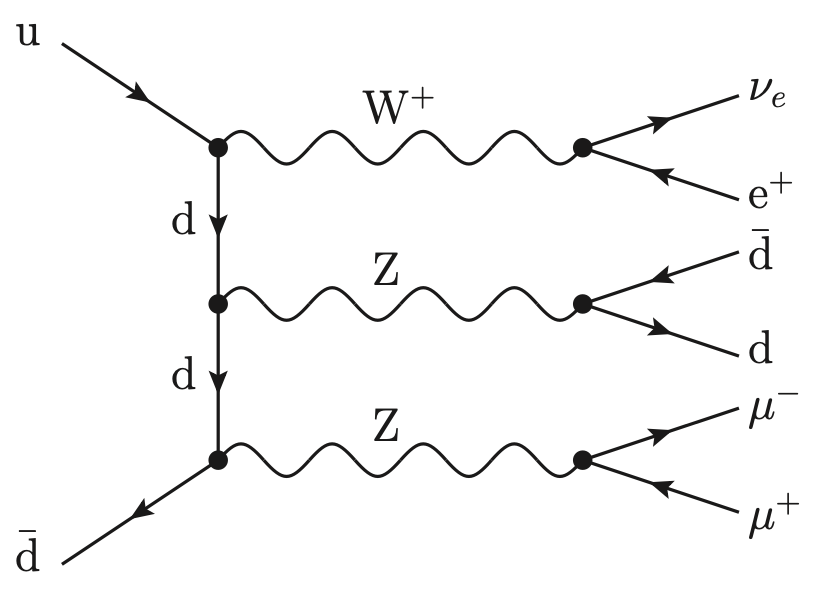}}
  \subfigure[\label{fig:a6_q2}]{\includegraphics[scale=0.45]{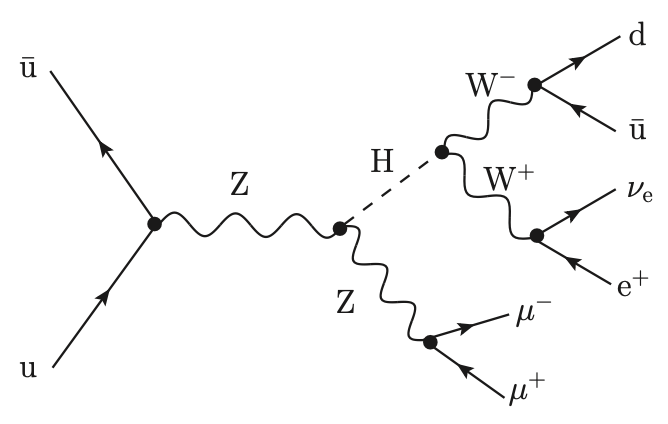}}
  \subfigure[\label{fig:a6_q3}]{\includegraphics[scale=0.46]{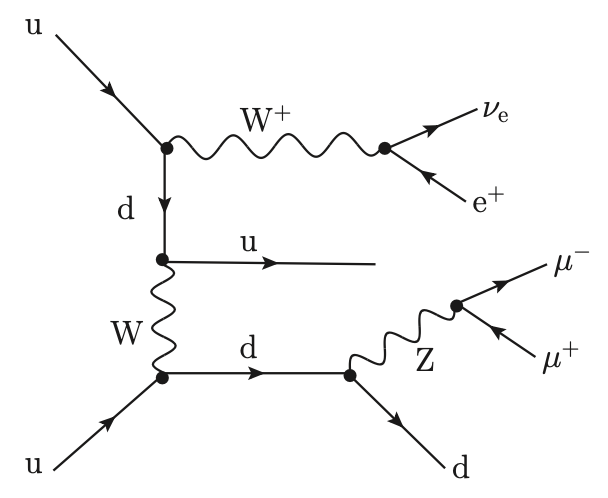}}
  \caption{Sample $\mathcal{O}(g^6)$ diagrams for quark-induced channels. Shown are topologies
    compatible with a tri-boson signature~[\ref{fig:a6_q1}], Higgs strahlung~[\ref{fig:a6_q2}],
    and a doubly-resonant contribution to the tri-boson background~[\ref{fig:a6_q3}].}\label{fig:a6_q}
\end{figure}
The bulk of the cross section arises from partonic channels that are compatible
with a tri-boson signal, namely
\beq\label{eq:lo1_qqb}
q_1\,\bar{q}_2\to\mu^+\mu^-\Pe^+\nu_\Pe\,q_3\,\bar{q}_4\,, \quad
\,q_i\in S_q=\{\Pu,\,\Pd,\,\Ps,\,\Pc\},\,
\eeq
where $q_3$ and $\bar{q}_4$ belong to the same quark generation and whose electric charge $Q$ is such
that $Q(q_3)+Q(\bar{q}_4)\in\{-1,\,0\}$. Predictions for these channels are dominated by genuine
tri-boson topologies [\reffi{fig:a6_q1}] or Higgs-strahlung diagrams [\reffi{fig:a6_q2}].
All remaining quark-induced
channels involving quarks $q_i\in S_q$ are not compatible with tri-boson production and can be summarised in the reactions
\beq\label{eq:lo1_qq}
q_1\,q_2\to\mu^+\mu^-\Pe^+\nu_\Pe\,q_3\,q_4\,,\qquad\bar{q}_1\,\bar{q}_2\to\mu^+\mu^-\Pe^+\nu_\Pe\,\bar{q}_3\,\bar{q}_4\,,
\quad \quad q_1\,\bar{q}^\prime_2\to\mu^+\mu^-\Pe^+\nu_\Pe\,q_3\,\bar{q}^\prime_4\,,
\eeq
where the notation $\bar{q}^\prime_i$ highlights that $q_1$ and
$\bar{q}^\prime_2$ as well as  $q_3$ and $\bar{q}^\prime_4$ belong to different
generations. These channels can be at most doubly resonant
and therefore represent a background to the tri-boson signature. Indeed they can
include for instance VBS-like topologies
or $t$-channel configurations, as the one illustrated in~\reffi{fig:a6_q3}, where the two quark lines
running from the initial to the final state exchange a $t$-channel vector boson. However, 
at the perturbative order we are considering, they are largely suppressed by the selection cuts
defining the signal region, in particular the cut
$M_{\Pj_1\Pj_2}<100\GeV$ in \refeq{eq:setup2}.

\begin{figure}
  \centering
  \subfigure[\label{fig:a6_a1}]{\includegraphics[scale=0.42]{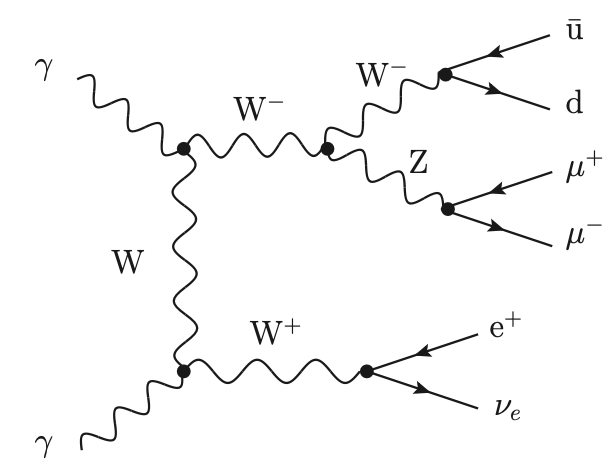}}
  \subfigure[\label{fig:a6_b1}]{\includegraphics[scale=0.28]{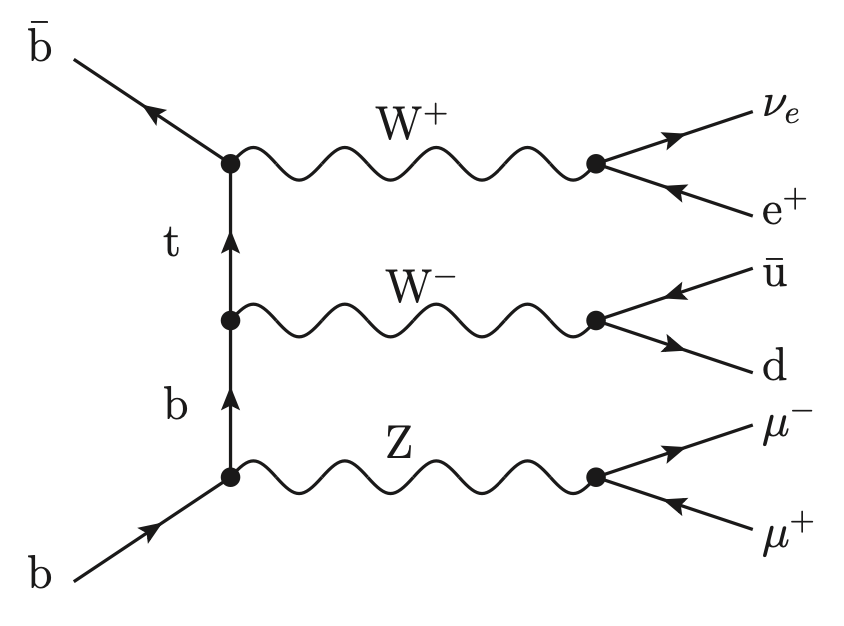}}
    \subfigure[\label{fig:a6_b2}]{\includegraphics[scale=0.28]{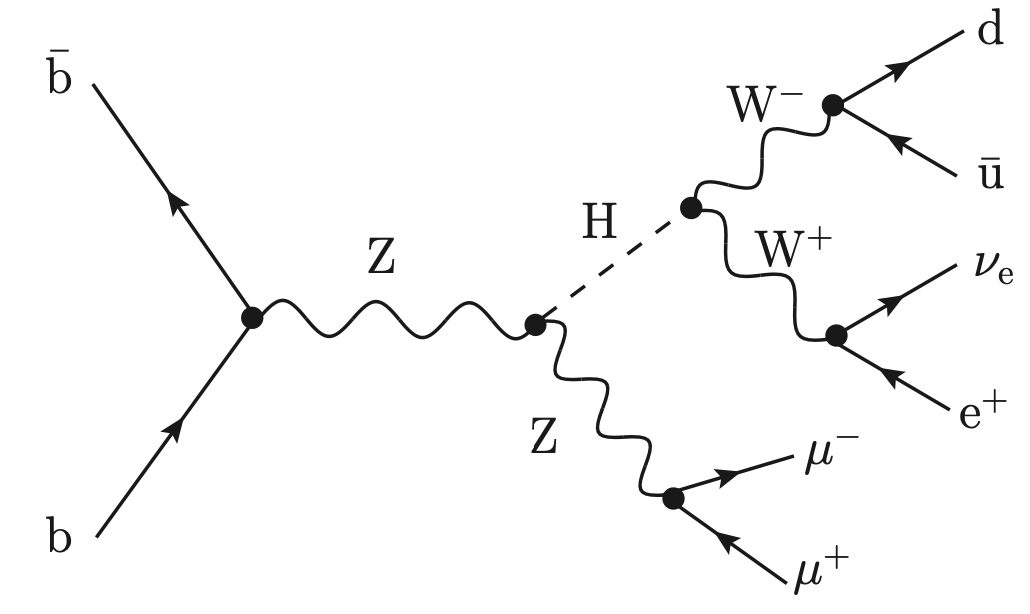}}
  \caption{Sample $\mathcal{O}(g^6)$ diagrams for $\gamma\gamma$- and $\Pb\bar{\Pb}$-induced channels. Shown are topologies
    compatible with a tri-boson signature for the $\gamma\gamma$ channel~[\ref{fig:a6_a1}] and the $\Pb\bar{\Pb}$ channel~[\ref{fig:a6_b1}].
    For the $\Pb\bar{\Pb}$ case a Higgs-strahlung contribution is also shown~[\ref{fig:a6_b2}].}\label{fig:a6_aa_b}
\end{figure}
At $\mathcal{O}(\alpha^6)$, two additional kinds of partonic channels are included as part of our
signal, namely the $\gamma\gamma$- and $\Pb\bar{\Pb}$-induced ones:
\begin{align}\label{eq:nlo1_aa}
  \gamma\,\gamma\,\to \mu^+\mu^-\Pe^+\nu_\Pe\,q_1\,\bar{q}_2\,,\qquad  \Pb\,\bar{\Pb}\to\mu^+\mu^-\Pe^+\nu_\Pe\,q_1\,\bar{q}_2\,.%\gamma\,.
\end{align}
Indeed, both of them contribute to tri-boson production with triply-resonant diagrams
like the ones shown in~\reffi{fig:a6_aa_b}, which partly compensate for the suppression from photon- and
bottom PDFs. In particular, we note that
the cross section for $\Pb\bar{\Pb}$-initiated channels is dominated by 
topologies involving a non-resonant $t$-channel top
quark~[\reffi{fig:a6_b1}] and Higgs-strahlung diagrams~[\reffi{fig:a6_b2}].

\begin{figure}\label{fig:a56}
  \centering
  \subfigure[\label{fig:a6_top}]{\includegraphics[scale=0.41]{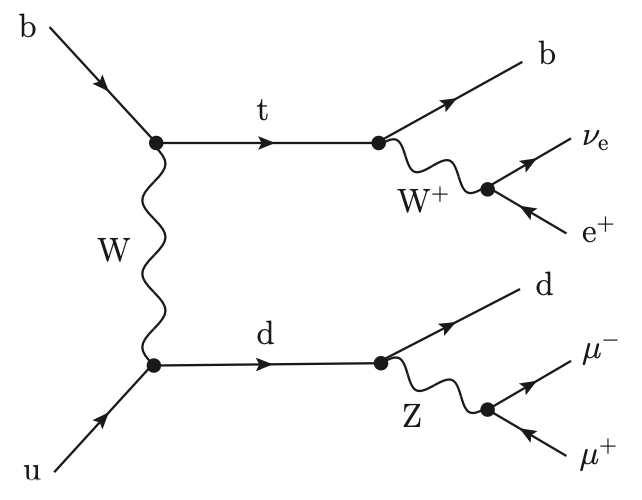}}
  \subfigure[\label{fig:a5_f1}]{\includegraphics[scale=0.41]{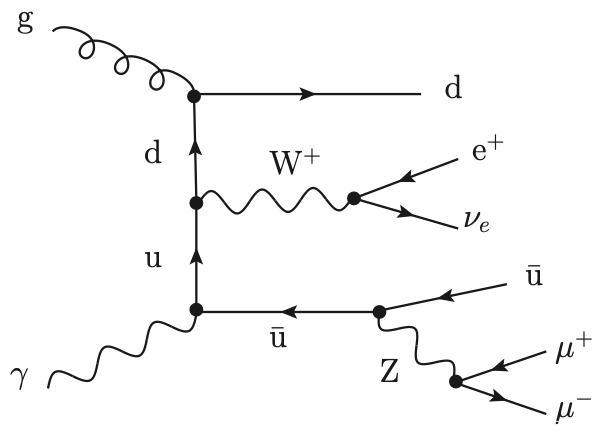}}
    \subfigure[\label{fig:a5_f2}]{\includegraphics[scale=0.40]{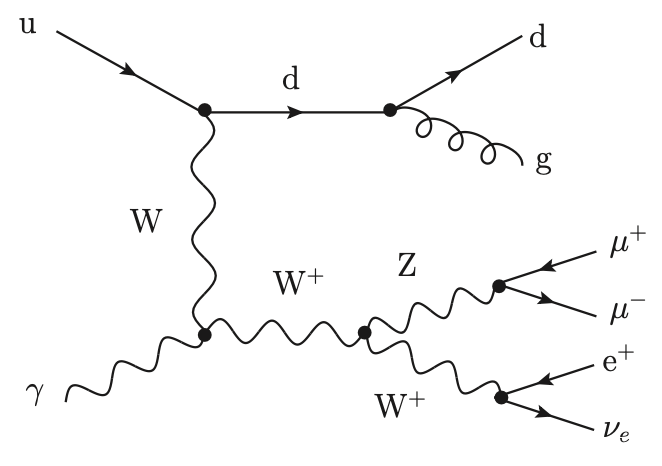}}
    \caption{Sample $\mathcal{O}(g^6)$ diagram for the $q\,\Pb$-induced channel with an $s$-channel
      top-quark resonance [\ref{fig:a6_top}] and sample $\mathcal{O}(g_{\rm s}g^5)$ diagrams
      for the $g\gamma$- and $\gamma q$-induced contributions [\ref{fig:a5_f1} and \ref{fig:a5_f2}].}\label{fig:a6_a5}
\end{figure}
To assess the impact of additional background sources to our signal region,
we also evaluate partonic processes involving bottom quarks in the final states,
which we separate and ultimately exclude in our final predictions with the assumption of a perfect $\Pb$-jet
veto.%
\footnote{These processes contribute to $\Pt\PZ\Pj$ production, which has been
  investigated in \citere{Denner:2022fhu}.}
They belong to the following two classes:
\begin{align}\label{eq:lo1_qb}
& q_1\,\Pb\to\mu^+\mu^-\Pe^+\nu_\Pe\,q_2\,\Pb\,,\qquad
  q_1\,\bar\Pb\to\mu^+\mu^-\Pe^+\nu_\Pe\,q_2\,\bar\Pb\,,\qquad \notag\\
& \bar q_1\,\Pb\to\mu^+\mu^-\Pe^+\nu_\Pe\,\bar q_2\,\Pb\,,\qquad
 \bar q_1\,\bar\Pb\to\mu^+\mu^-\Pe^+\nu_\Pe\,\bar q_2\,\bar\Pb\,,\qquad \notag\\
& q_1\,\bar{q}_2\to\mu^+\mu^-\Pe^+\nu_\Pe\,\Pb\,\bar{\Pb}\,.
\end{align}
The contributions  in~\refeq{eq:lo1_qb} with a single $\Pb$
or $\bar\Pb$ quark in the initial state can
be at most doubly resonant in the vector bosons but are enhanced by the appearance of
a top-quark resonance, as in the
diagram shown in~\reffi{fig:a6_top}. For the contributions in~\refeq{eq:lo1_qb}
with a final-state $\Pb\bar{\Pb}$ pair, top-quark-resonant topologies are accompanied by tri-boson
diagrams, where the bottom quarks result from  the hadronic decay of a $\PZ$ boson.

\subsubsection{Contributions to $\mathcal{O}(\alphas \alpha^5)$}
\label{sec:lo2}

The $\mathcal{O}(\alphas \alpha^5)$ can only receive contributions from topologies which are at most
doubly resonant in the vector bosons and therefore part of the tri-boson background. %Nevertheless,
%since the EW corrections to this LO contribution can not be
Two different combinations of
 amplitudes enter this perturbative order: interferences of
$\mathcal{O}(g^6)$ and $\mathcal{O}(\gs^2 g^4)$ amplitudes and squares of $\mathcal{O}(\gs g^5)$ amplitudes.

Contributions involving two initial-state quarks can only appear as
interference terms. The latter vanish owing to colour algebra,
except for the case where $t$-channel diagrams interfere with $u$- and/or $s$-channel ones. This
explains why no contribution involving bottom quarks, either as initial or final states, appears at
this order.

At $\mathcal{O}(\gs g^5)$  only amplitudes with one external gluon and one external photon are
possible. That allows for new partonic channels, \ie the $\Pg\gamma$-,
$\gamma q$-, and $\gamma \bar q$-induced ones:
\begin{align}\label{eq:lo2_ag}
 \Pg\,\gamma\,\to\mu^+\mu^-\Pe^+\nu_\Pe\,q_1\, \bar{q}_2\,,\qquad
 \gamma\,q_1\to\mu^+\mu^-\Pe^+\nu_\Pe\,q_2\,\Pg\,, \qquad
 \gamma\,\bar q_1\to\mu^+\mu^-\Pe^+\nu_\Pe\,\bar{q}_2\,\Pg\,.
\end{align}
No $\gamma\Pb$- or $\gamma\bar{\Pb}$-induced channels contribute to our results. The former channel is simply
forbidden by charge conservation. The second one can just be constructed by
including a top quark in the final state  at this perturbative order.
However, since the top quark is treated as a resonance throughout our
calculation, it can not be part of the final state, so that all $\gamma\bar{\Pb}$-induced diagrams 
are discarded.
Some illustrative diagrams for the $\Pg\gamma$ and $\gamma q$ partonic
processes are reported in~\reffis{fig:a5_f1} and~\ref{fig:a5_f2}, respectively. It is worth noting that the bulk
of the $\gamma q/\gamma\bar{q}$ cross section originates from diagrams involving a
$q\to\Pg q$ or $\bar{q}\to\Pg\bar{q}$ final-state splitting as in~\reffi{fig:a5_f2}.
Indeed, this final-state quark--gluon pair can easily be misidentified as a pair of jets from a would-be
hadronically-decaying vector boson, so that the event can pass the tri-boson selection cuts. That is in contrast
to what happens for the $\Pg\gamma$ channel, where the two final-state quarks are connected by a $t$-channel
fermion line and on average have a larger invariant mass, which renders it less likely for the events to
pass the cuts. This fact is important to understand the relative size of these
contributions observed in~\refse{sec:integrated}.

\subsubsection{Contributions to $\mathcal{O}(\alphas^2 \alpha^4)$}
\label{sec:lo3}

At this perturbative order only squares of $\mathcal{O}{( \gs^2 g^4 )}$ amplitudes contribute,
which are again at most doubly resonant and comprise either an
internal gluon propagator or two external gluon lines.

\begin{figure}
  \centering
  \subfigure[\label{fig:a4_qq}]{\includegraphics[scale=0.27]{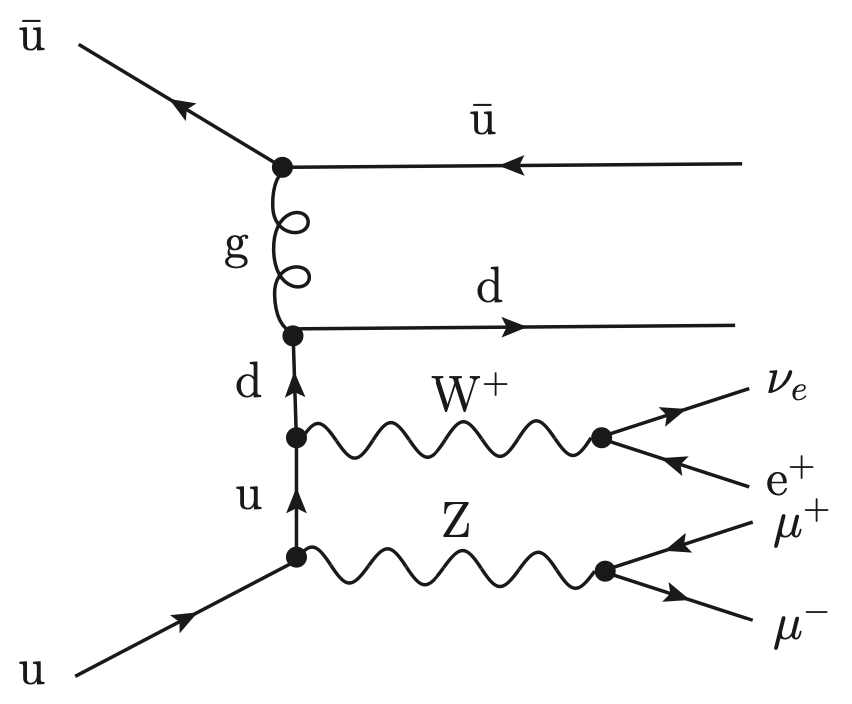}}
%  \hspace{0.1\textwidth}
  \subfigure[\label{fig:a4_bb}]{\includegraphics[scale=0.41]{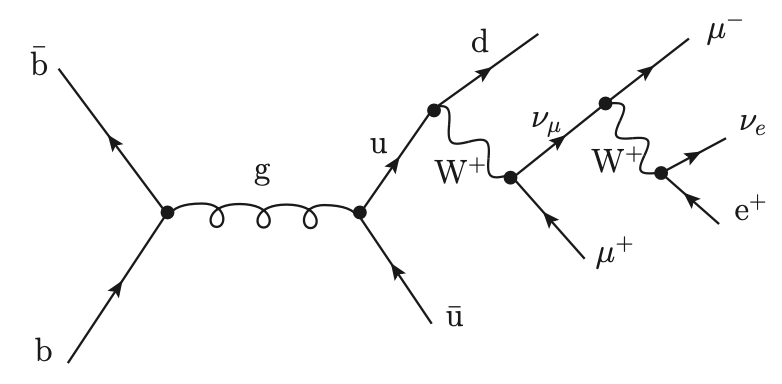}}
  \subfigure[\label{fig:a4_bb_fs}]{\includegraphics[scale=0.41]{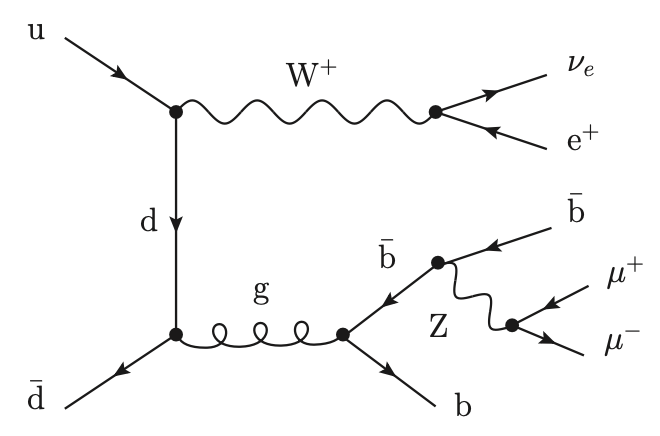}}
  \caption{Sample $\mathcal{O}(g_{\rm s}^2 g^4)$ diagrams for light--quark-induced channels~[\ref{fig:a4_qq}],
    bottom-induced channels~[\ref{fig:a4_bb}] and for channels with
    two bottom quarks in the final state~[\ref{fig:a4_bb_fs}].}\label{fig:a4_q}
\end{figure}
Among contributions with no external gluons, the dominant ones involve
$q\bar{q}^\prime$-, $qq^\prime$-, and
$\bar{q}\bar{q}^\prime$-induced channels. These can only proceed via the exchange of an internal
$t$-channel gluon, as exemplified in~\reffi{fig:a4_qq}, while the $\Pb\bar{\Pb}$-induced ones, which are
largely suppressed by the bottom PDFs, only
include diagrams with an $s$-channel gluon, as shown
in~\reffi{fig:a4_bb}. As for $\mathcal{O}(\alpha^6)$, a sizeable
fraction of the cross section arises from channels with final-state
bottom quarks [\reffi{fig:a4_bb_fs}], which at this order are not
enhanced by the presence of a top-quark resonance.

\begin{figure}
  \centering
  \subfigure[\label{fig:a4_gg}]{\includegraphics[scale=0.48]{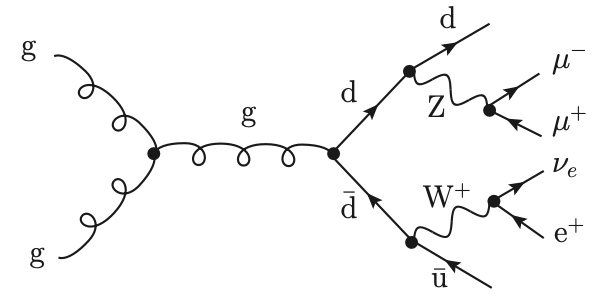}}
  \subfigure[\label{fig:a4_qg}]{\includegraphics[scale=0.46]{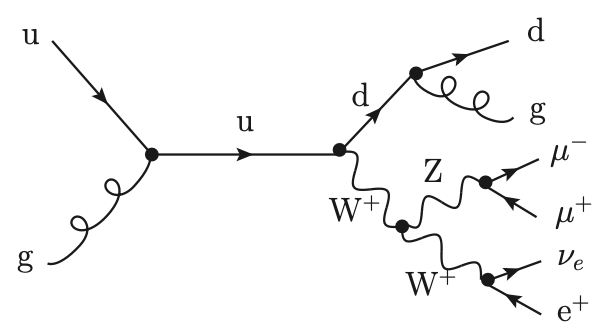}}
  \subfigure[\label{fig:a4_gg_fs}]{\includegraphics[scale=0.42]{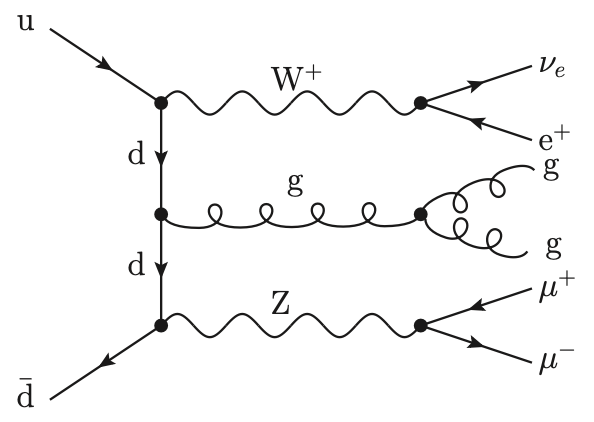}}
  \caption{Sample $\mathcal{O}(g_{\rm s}^2 g^4)$ diagrams involving two external gluons. Shown are
    topologies with two initial-state gluons~[\ref{fig:a4_gg}], one initial-state gluon~[\ref{fig:a4_qg}],
    and two final-state gluons~[\ref{fig:a4_gg_fs}].}\label{fig:a4_g}
\end{figure}
Channels including external gluons provide by far the dominant contributions across all perturbative orders considered
in this paper and are described by the reactions
\begin{align}
  &\Pg\,\Pg\,\to  \mu^+\mu^-\Pe^+\nu_\Pe\,q_1\bar{q}_2\,,\qquad  
    q_1\,\bar{q}_2\,\to\mu^+\mu^-\Pe^+\nu_\Pe\,\Pg\,\Pg\,,\notag\\
\label{eq:lo3_gg}
 & q_1\,\Pg\,\to\mu^+\mu^-\Pe^+\nu_\Pe\,q_2\,\Pg\,,\qquad
  \bar{q}_1\,\Pg\,\to\mu^+\mu^-\Pe^+\nu_\Pe\,\bar{q}_2\,\Pg\,.
\end{align}
Although $\Pg\Pg$-induced channels are expected to be enhanced by the gluon PDFs, they appear to
be suppressed by the definition of the tri-boson fiducial region,
where the final-state QCD partons are required to have an invariant mass compatible with a hadronically-decaying
vector boson [see~\reffi{fig:a4_gg}]. Indeed, the invariant mass of the two final-state quarks
connected by internal fermion lines is expected to have higher values on average.
This is not the case for the $q\Pg/\bar{q}\Pg$-induced channels or the $q\bar{q}$ ones
with two final-state gluons, shown in~\reffi{fig:a4_qg}
and~\reffi{fig:a4_gg_fs}, respectively. In each case, the $q\Pg$ and
$\Pg\Pg$~pair arising from a QCD-singular splitting 
can have an invariant mass of the order of $\MW$ and therefore can be
reconstructed as a vector boson.

Note that channels involving bottom quarks can not be obtained from \refeq{eq:lo3_gg}
with the replacement $(q_1,\bar{q}_2)\to(\Pb,\bar{\Pb})$ or $(q_1,q_2)\to(\Pb,\Pb)$,
since charge conservation forces $Q(q_1)+Q(\bar{q}_2)=Q(q_1)-Q(q_2)=+1$.

\subsection{Next-to-leading-order corrections}
\label{sec:nexttoleadingorder}

In our calculation, we consider the NLO EW and QCD corrections to the $\mathcal{O}{( \alpha^6 )}$ cross section, 
which accounts for the signal. We also evaluate  NLO EW corrections to $\mathcal{O}{\left( \alphas \alpha^5 \right)}$, since
they can not be disentangled from NLO QCD corrections to $\mathcal{O}{\left( \alpha^6 \right)}$ in an infrared(IR)-safe manner.
Nevertheless, NLO corrections of orders 2 and more in $\alphas$ are
not computed, \ie we do consider neither NLO QCD corrections to the 
$\mathcal{O}{\left( \alphas \alpha^5 \right)}$ cross section nor NLO QCD and EW corrections to the $\mathcal{O}{\left( \alphas^2 \alpha^4 \right)}$ 
cross section. The latter two would  represent corrections to the
background and can safely be dropped
without breaking gauge invariance or spoiling the IR safety of our calculation.

Both QCD and QED singularities of soft and collinear origin that plague the real
contributions are treated using the dipole subtraction formalism
\cite{Catani:1996vz,Dittmaier:1999mb,Catani:2002hc,Dittmaier:2008md}.
The initial-state collinear singularities are absorbed in the PDFs in the $\MSbar$ factorisation scheme. 

Throughout our calculation, the complex-mass scheme for all unstable particles is used
\cite{Denner:1999gp,Denner:2005fg,Denner:2006ic,Denner:2019vbn}
resulting in complex input values for the EW boson masses,
the top-quark mass, and the EW mixing angle,
\beq
\mu_B^2 = M^2_B-\ri\Gamma_B M_B\quad (B\in{\PW,\PZ,\PH})\,,\qquad 
\mu_\Pt^2 = \Mt^2-\ri\Gt \Mt\,,\qquad 
\cos^2\theta_{\rw} = \frac{\mu_{\PW}^2}{\mu_{\PZ}^2}\,.
\eeq

\subsubsection{Contributions to $\mathcal{O}(\alpha^7)$}
\label{sec:nlo1}

\begin{figure}
        \centering
        \subfigure[\label{fig:a7_au_ISR}]{\includegraphics[scale=0.48]{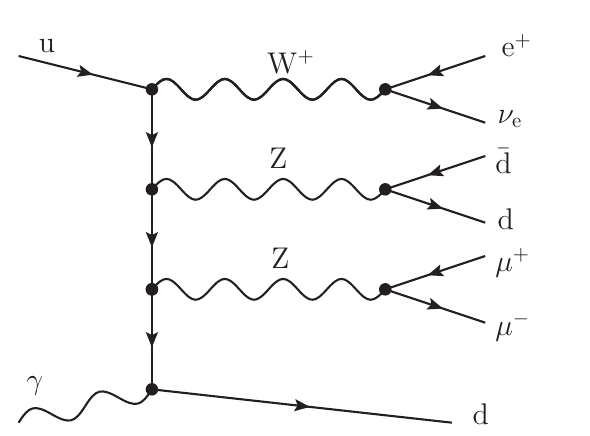}}
        \subfigure[\label{fig:a7_aa_afroml}]{\includegraphics[scale=0.52]{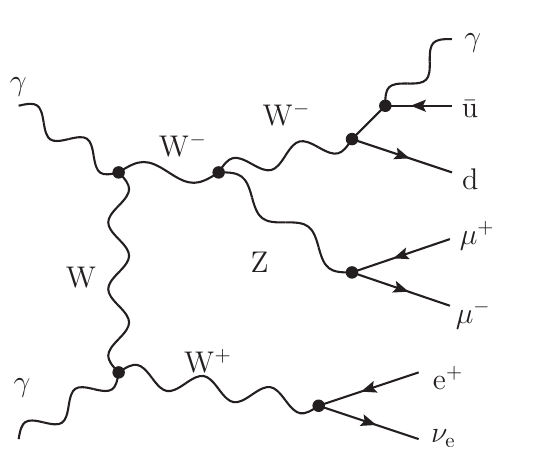}}
        \subfigure[\label{fig:a7_aa}]{\includegraphics[scale=0.52]{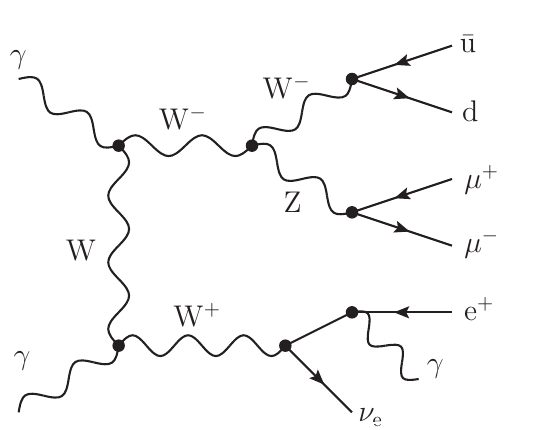}}
        \subfigure[\label{fig:a7_qqx_aISR}]{\includegraphics[scale=0.48]{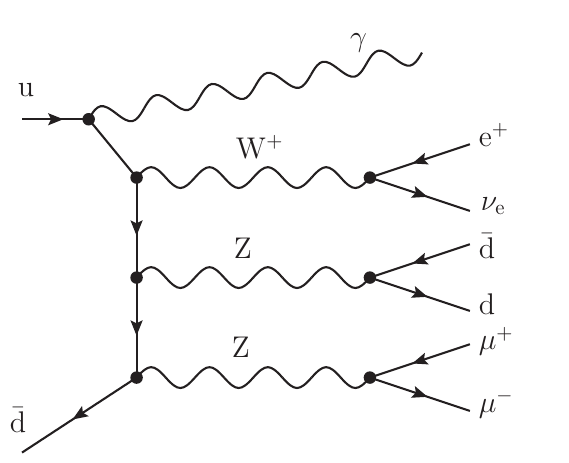}}
        \subfigure[\label{fig:a7_udx_aFSR}]{\includegraphics[scale=0.52]{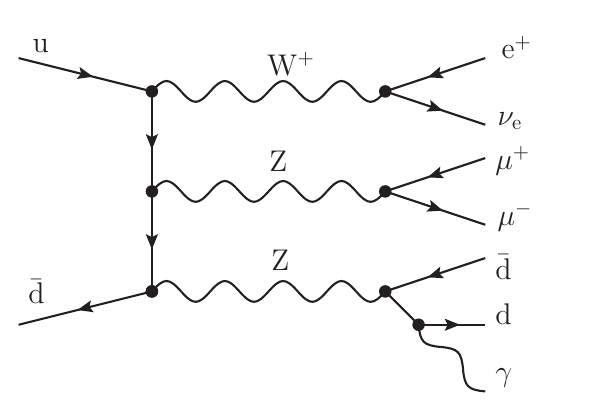}}
        \caption{Sample $\mathcal{O}(g^7)$ diagrams involving external
          photons. We show topologies with one and two initial-state
          photons [\ref{fig:a7_au_ISR}, \ref{fig:a7_aa_afroml}, and
          \reffi{fig:a7_aa}] 
          as well as with initial-state quark--antiquark pairs
          [\ref{fig:a7_qqx_aISR} and \ref{fig:a7_udx_aFSR}].}
        \label{fig:a7_a}
\end{figure}

NLO EW corrections to the $\mathcal{O}(\alpha^6)$ comprise two kinds of contributions. The first class
is represented by single-photon-induced terms, which are purely real
corrections to EW-mediated quark-induced channels and described by the reactions
\beq\label{eq:nlo1_qa}
\gamma\,q_1\to\mu^+\mu^-\Pe^+\nu_\Pe\,q_2\,q_3\bar{q}_4\,,\qquad
\gamma\,\bar{q}_1\to\mu^+\mu^-\Pe^+\nu_\Pe\,\bar{q}_2\,q_3\bar{q}_4\,.
\eeq
The presence of a photon in the initial state only introduces an initial-state singularity,
which is absorbed via PDF renormalisation. A representative diagram for this class is
shown in~\reffi{fig:a7_au_ISR}.

The second class of NLO EW contributions includes both virtual and real corrections to the partonic
channels entering at $\mathcal{O}(\alpha^6)$. The real corrections to the $\gamma\gamma$ channel,
namely
\beq\label{eq:nlo1_aa2}
\gamma\,\gamma\to\mu^+\mu^-\Pe^+\nu_\Pe\,q_1\,\bar{q}_2\,\gamma\,,
\eeq
have a larger number of singular regions compared to the case in \refeq{eq:nlo1_qa},
since the real final-state photon can be radiated both by final-state quarks and by charged leptons,
as illustrated in~\reffis{fig:a7_aa_afroml} and \ref{fig:a7_aa}, respectively.
However, the real channels with two quarks in the initial state,
\begin{align}
&  q_1\,\bar{q}_2\to
\mu^+\mu^-\Pe^+\nu_\Pe\,q_3\bar{q}_4\,\gamma\,,\qquad\!\!
&& q_1\,q_2\to\mu^+\mu^-\Pe^+\nu_\Pe\,q_3\,q_4\,\gamma\,,\qquad\!\! 
&&\bar{q}_1\,\bar{q}_2\to\mu^+\mu^-\Pe^+\nu_\Pe\,\bar{q}_3\,\bar{q}_4\,\gamma\,,\notag\\
&  q_1\,\bar{q}^\prime_2\to\mu^+\mu^-\Pe^+\nu_\Pe\,q_3\,\bar{q}^\prime_4\,\gamma\,,\qquad\!\!
&& \Pb_1\,\bar{\Pb}_2\to\mu^+\mu^-\Pe^+\nu_\Pe\,q_1\bar{q}_2\,\gamma\,,
\label{eq:nlo1_qq}
\end{align}
are the ones with the richest IR structure, since they can contain both
initial- and final-state collinear singularities (see
\reffis{fig:a7_qqx_aISR} and \ref{fig:a7_udx_aFSR} for exemplary diagrams).
For this reason, the subtraction procedure and the numerical integration of
these terms are among the most computationally intensive in our calculation.

\begin{figure}
        \centering
        \subfigure[\label{fig:a7_qq_virt}]{\includegraphics[scale=0.50,width=5.2cm,height=5cm]{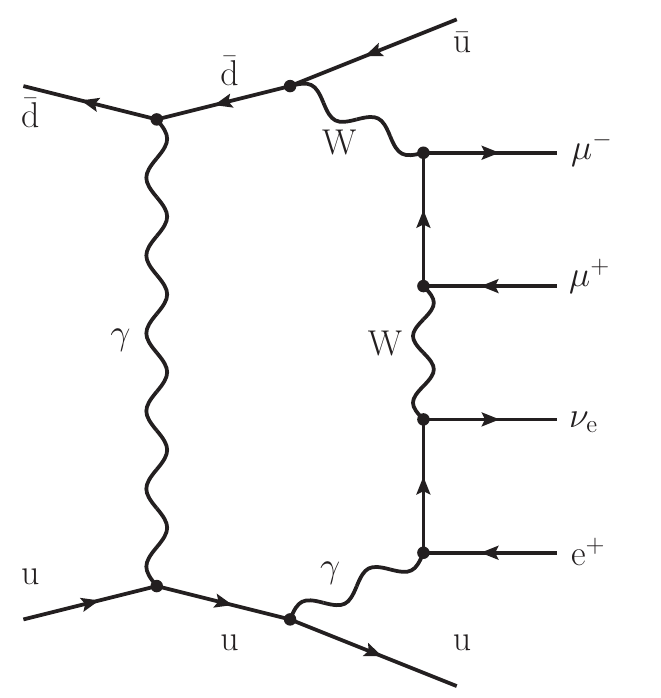}}
        \subfigure[\label{fig:a7_aa_virt}]{\includegraphics[scale=0.50,width=5.2cm,height=5cm]{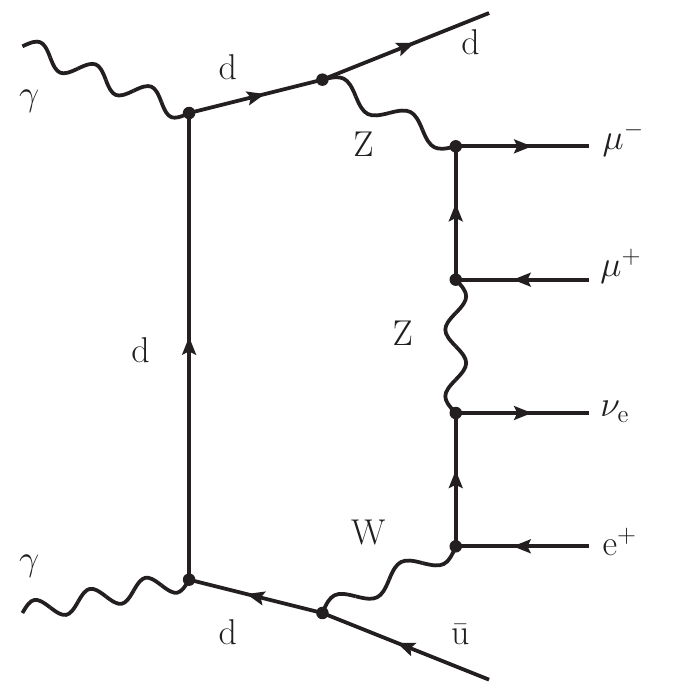}}
        \caption{One-loop diagrams of $\mathcal{O}(g^8)$ involving
          tensor integrals of rank 4 [\ref{fig:a7_qq_virt}] and rank 5
          [\ref{fig:a7_aa_virt}].}\label{fig:a7_virt}
\end{figure}

For this second class of partonic processes the evaluation of the virtual contributions, whose amplitudes
result from the interference of $\mathcal{O}(g^6)$ diagrams with $\mathcal{O}(g^8)$ loop ones, is also
a computationally expensive part of the calculation. The complexity 
arises both from the number of loop diagrams to be accounted for, which can become significantly
large for fully EW diagrams, and from the evaluation of
the loop integrals, which can involve up to $8$-point functions with tensor rank
$4$~[\reffi{fig:a7_qq_virt}] and $5$~[\reffi{fig:a7_aa_virt}] for the quark- and $\gamma\gamma$-induced channels, respectively.

\subsubsection{Contributions to $\mathcal{O}(\alphas\alpha^6)$}
\label{sec:nlo2}

This perturbative order receives contributions both from NLO QCD corrections to
$\mathcal{O}(\alpha^6)$ and from NLO EW corrections to $\mathcal{O}(\alphas\alpha^5)$.

\begin{figure}
        \centering
        \subfigure[t][\label{fig:as1a6_qq_faf_lower}]{\includegraphics[width=0.26\linewidth]{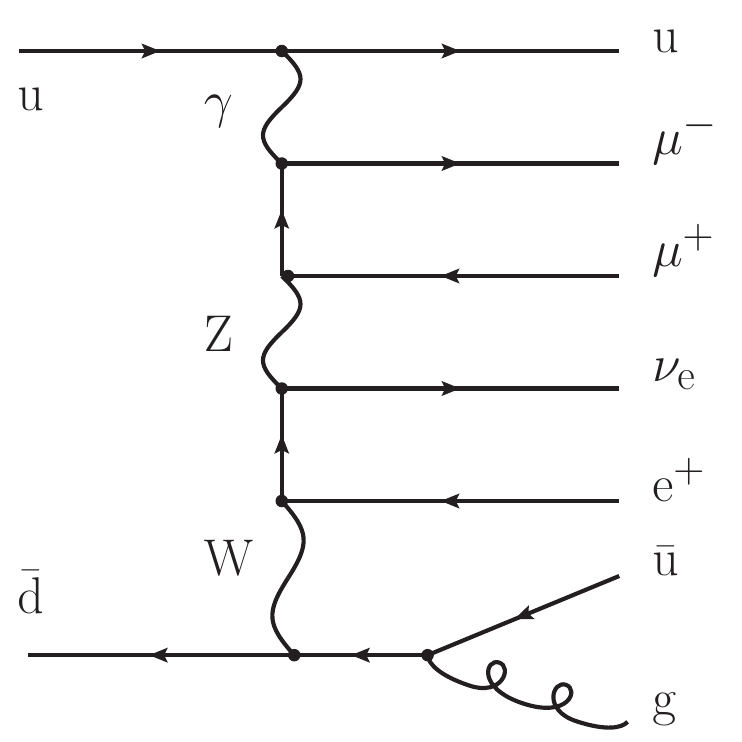}}\qquad
        \subfigure[t][\label{fig:as1a6_qq_faf}]{\includegraphics[width=0.26\linewidth]{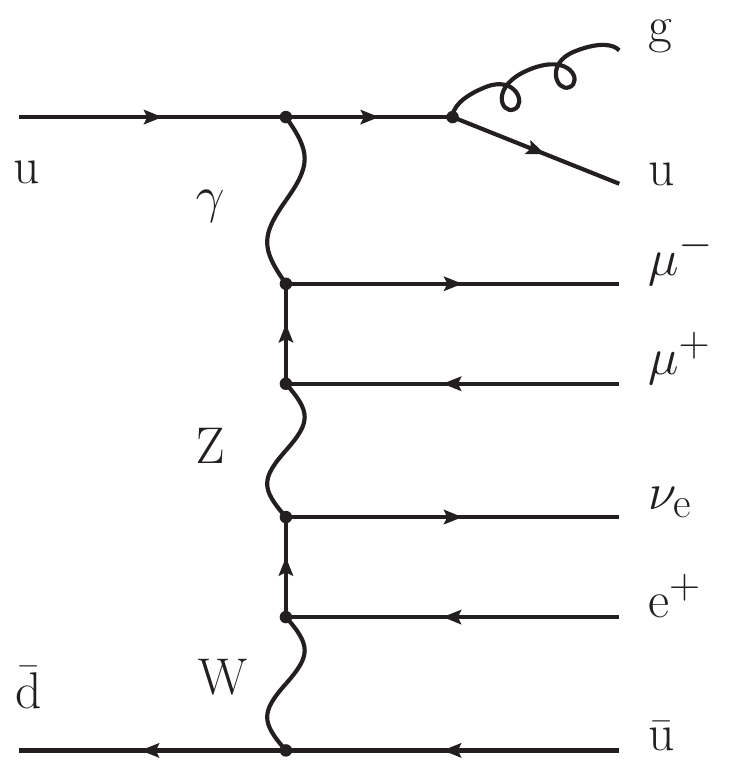}} \\ 
        \subfigure[t][\label{fig:as1a6_qq_aff}]{\includegraphics[width=0.34\linewidth]{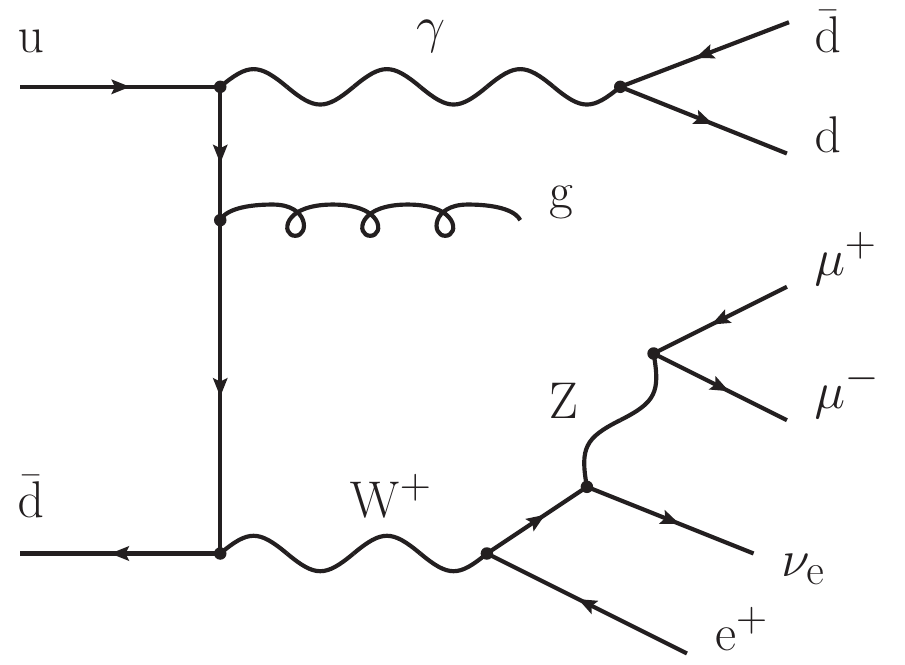}} \qquad
        \subfigure[t][\label{fig:as1a6_gq}]{\includegraphics[width=0.38\linewidth]{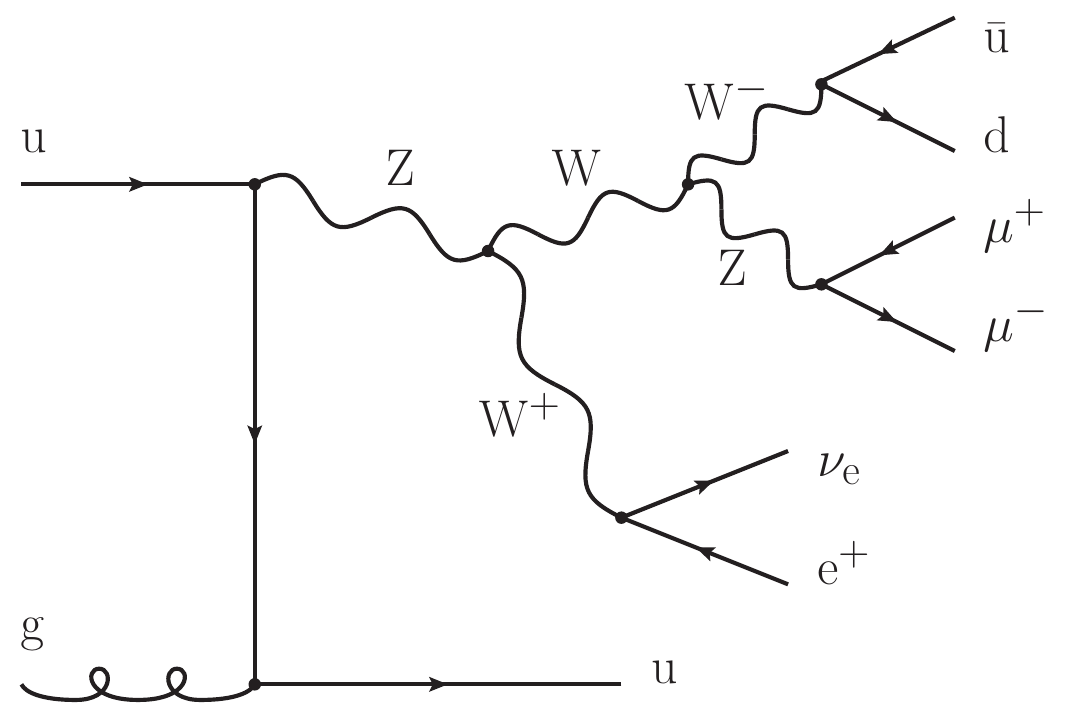}}
        \caption{Exemplary diagrams of $\mathcal{O}(g_{\rm s}g^6)$ with one external gluon 
          involving an initial-state singularity from a $q\to q\gamma$ splitting
          [\ref{fig:as1a6_qq_faf_lower}], a final-state singularity from
          a $\gamma\to q\bar{q}$ splitting [\ref{fig:as1a6_qq_aff}]
          and without QED singularities
          [\ref{fig:as1a6_qq_faf} and \ref{fig:as1a6_gq}].}
        \label{fig:as1a6_g}
\end{figure}

As far as the real contributions are concerned, one can still distinguish the two kinds
of corrections. We start considering real QCD corrections to $\mathcal{O}(\alpha^6)$, which result
from squaring amplitudes of $\mathcal{O}(\gs g^6)$.
These comprise partonic channels at $\mathcal{O}(\alpha^6)$ with an additional final-state gluon,
\begin{align}
\!\!& q_1\,\bar{q}_2\to  \mu^+\mu^-\Pe^+\nu_\Pe\,q_3\bar{q}_4\,\Pg\,,\label{eq:nlo2_qqb}\\
\!\!& q_1\,q_2\to \mu^+\mu^-\Pe^+\nu_\Pe\,q_3\,q_4\,\Pg\,,\quad 
&& \bar{q}_1\,\bar{q}_2\to\mu^+\mu^-\Pe^+\nu_\Pe\,\bar{q}_3\,\bar{q}_4\,\Pg\,, \quad 
&& q_1\,\bar{q}^\prime_2\to\mu^+\mu^-\Pe^+\nu_\Pe\,q_3\,\bar{q}^\prime_4\,\Pg\,, \label{eq:nlo2_qq}\\
\!\!& \gamma\,\gamma\to\mu^+\mu^-\Pe^+\nu_\Pe\,q_1\,\bar{q}_2\,\Pg\,, \quad
&& \Pb\,\bar{\Pb}\to\mu^+\mu^-\Pe^+\nu_\Pe\,q_1\,\bar{q}_2\,\Pg\,,\label{eq:nlo2_aa}
\end{align}
as well as new partonic channels involving an initial-state gluon,
\begin{align}\label{eq:nlo2_gq}
  q_1\,\Pg\,\to \mu^+\mu^-\Pe^+\nu_\Pe\,q_2\,q_3\,\bar{q}_4\,,\qquad
  \bar{q}_1\,\Pg\,\to \mu^+\mu^-\Pe^+\nu_\Pe\,\bar{q}_2\,q_3\,\bar{q}_4\,.
\end{align}
We note that, although the terms in \refeqs{eq:nlo2_qqb}--\eqref{eq:nlo2_gq} represent NLO QCD
corrections, in general
both QCD and QED Catani--Seymour (CS) dipoles are needed to properly account for all
IR singularities. This is indeed true for all channels with at least one initial-state quark,
while the $\gamma\gamma$- and $\Pb\bar{\Pb}$-induced channels in \refeq{eq:nlo2_aa}
have a simpler structure and
only suffer from QCD divergences.
In \reffi{fig:as1a6_qq_faf_lower} a diagram with an initial-state collinear QED singularity
from a $q\to q\gamma$ splitting is shown as an example. Final-state collinear QED divergences
can also appear and are separately discussed in the following.

Among the real channels with at least one initial-state quark $q$, the ones in~\refeq{eq:nlo2_qq}, 
which are not compatible with a tri-boson signal, have the simpler IR
structure, since
QED CS dipoles are required only for initial-state singularities 
besides the QCD dipoles.  However, the behaviour of the
NLO corrections is particularly interesting for these partonic processes.
Indeed, despite the high suppression of the $\mathcal{O}(\alpha^6)$ induced by the selection cuts, they
receive large and positive corrections at
$\mathcal{O}(\alphas\alpha^6)$, as it is confirmed
by the results in~\refse{sec:integrated}. This can be understood as
follows: While all contributions 
with a Born-like kinematics are largely cut away from our signal
region by the cuts in~\refeqs{eq:setup1}
and~\eqref{eq:setup2}, this is not the case
for the real contributions, where the additional gluon radiation
allows to elude the di-jet invariant-mass cut of~\refeq{eq:setup2}
whenever a quark--gluon pair is misidentified as the hadronic decay
of a would-be vector boson. These contributions obviously miss the
enhancement of one vector-boson resonance. On the other hand, they involve a $t$-channel
vector boson coupled to an incoming and outgoing quark as in
\reffi{fig:as1a6_qq_faf}, which is enhanced at high energies similar
to VBS topologies \cite{Denner:2024ufg}. 

The remaining channels with at least an initial-state quark $q$ are
the tri-boson-compati\-ble quark-induced ones in \refeq{eq:nlo2_qqb}
and the quark--gluon-initiated ones in \refeq{eq:nlo2_gq}, for which
diagrams are shown in \reffis{fig:as1a6_qq_faf_lower}--\ref{fig:as1a6_qq_aff} and
\reffi{fig:as1a6_gq}, respectively.  Both of them admit a virtual photon with a
subsequent splitting into a $\Pq\bar{\Pq}$ pair in the final state,
\ie via $\gamma^*\,\to\,q\bar{q}$ [see \reffi{fig:as1a6_qq_aff}], which renders the treatment of IR
QED singularities more involved. In the limit of vanishing virtuality
of the photon  the real amplitude develops a collinear singularity. The collinear
quark pair is clustered into a single jet and the singular event mapped into a Born-like
topology. By virtue of the KLN theorem~\cite{Kinoshita:1962ur,PhysRev.133.B1549},
this singularity would be cancelled by a proper inclusion
of virtual EW corrections to the corresponding $\mathcal{O}(\alpha^6)$ channels, which in our
case would be
\begin{align}
  q_1\,\bar{q}_2\to  \mu^+\mu^-\Pe^+\nu_\Pe\,\gamma\,\Pg\,,\qquad  
q_1\,\Pg\,\to \mu^+\mu^-\Pe^+\nu_\Pe\,q_2\,\gamma\,,\qquad  
\bar{q}_1\,\Pg\,\to \mu^+\mu^-\Pe^+\nu_\Pe\,\bar{q}_2\,\gamma\,.
\end{align}
Nevertheless, these channels are not part of our process definition, since
the photon is not treated as a jet.
To deal with this unsubtracted divergences, we employ the photon--to--jet conversion function introduced
in~\citere{Denner:2019zfp}. In this approach, also employed in $\PW\PZ$ scattering \cite{Denner:2019tmn},
  the singularity is absorbed into the conversion function together with 
non-perturbative contributions, which arise when integrating over the photon virtuality down to mass
scales of the light hadrons.

We move on to consider real NLO EW corrections to the $\mathcal{O}(\alphas\alpha^5)$.
These corrections can be obtained in two different
ways. A first contribution arises when interfering $\mathcal{O}(g^7)$ and $\mathcal{O}(\gs^2 g^5)$
amplitudes involving a real-photon emission. The $\Pb\bar{\Pb}$- and $\gamma\gamma$-induced channels
receive NLO QCD corrections, as discussed above, but no
NLO EW corrections, since the corresponding $\mathcal{O}(\as \alpha^5)$ vanishes owing to colour algebra.
%Only partonic channels with a non-zero $\mathcal{O}(\as \alpha^5)$ enter this order.
For this class of real amplitudes, an internal gluon could potentially cause an initial-state
singularity from a collinear $q\to q \Pg$ splitting or a final-state one from a $\Pg\to q\bar{q}$ splitting.
Nevertheless, since our signal region requires at least two jets (see \refse{sec:numresults}) and the
photon can not be promoted to a jet, these singular regions are cut away. This means that QED CS dipoles
are sufficient to subtract all the IR singularities of these real terms.

A second contribution includes squared $\mathcal{O}(\gs g^6)$ amplitudes, which correspond
to the real reactions:
\begin{align}\label{eq:nlo2_ga}
\Pg\,\gamma\,\to\mu^+\mu^-\Pe^+\nu_\Pe\,q_1\,\bar{q}_2\,\gamma\,,\qquad
\gamma\,q_1\,\to \mu^+\mu^-\Pe^+\nu_\Pe\,q_2\,\Pg\,\gamma\,,\qquad
\gamma\,\bar{q}_1\,\to \mu^+\mu^-\Pe^+\nu_\Pe\,\bar{q}_2\,\Pg\,\gamma\,.
\end{align}
For the same reasons discussed above, our fiducial cuts prevent any QCD singularity
in the $\Pg\,\gamma$-induced channel, which only suffers from
soft and collinear photon singularities. However, the $\gamma\,q$- and
$\gamma\,\bar{q}$-induced ones
require a special treatment. Whenever a soft gluon is isolated, the definition of
our fiducial region removes these singular
configurations, since they would lead to a mono-jet signature. The same occurs for gluons collinear to the initial/final state
parton $q_1/q_2$. However, within the dressing procedure, any QCD parton, and so also a soft gluon, can be
recombined with a hard photon.
This recombination step would lead to a two-jet event  with a soft-gluon singularity,
which should be compensated by
the virtual QCD correction to the partonic channel
\begin{align}
 \gamma\,q_1\,\to \mu^+\mu^-\Pe^+\nu_\Pe\,q_2\,\gamma\,,\qquad
 \gamma\,\bar{q}_1\,\to \mu^+\mu^-\Pe^+\nu_\Pe\,\bar{q}_2\,\gamma\,.
\end{align}
Again, the latter channels are not part of our LO process definition, since photons are never considered as jets.
Following the approach of \citeres{Denner:2014ina,Denner:2010ia,Denner:2009gj},
the unsubtracted singularity is removed by discarding all events with a jet that
arises from the recombination of a parton $p$ (a quark, an antiquark,
or a gluon) of energy $E_p$
with a photon of energy $E_\gamma$ having an energy fraction \hbox{$z_\gamma = E_\gamma/(E_\gamma+E_p)$}
that exceeds a cut value $z^{\rm cut}_\gamma$. We have used the standard value $z^{\rm cut}_\gamma=0.7$ throughout our
calculation. This democratic treatment of partons to remove jets containing hard photons is not IR safe. Indeed,
the cancellation of singularities from the virtual photonic
corrections to the channels $\gamma\,q_1\,\to
\mu^+\mu^-\Pe^+\nu_\Pe\,q_2\,\Pg$ and $\gamma\,\bar{q}_1\,\to
\mu^+\mu^-\Pe^+\nu_\Pe\,\bar{q}_2\,\Pg$ is spoilt
by cutting away real QED events whose collinear pair $(q_2,\gamma)$ has $z_\gamma>z^{\rm cut}_\gamma$.
This artificially introduced divergence is absorbed into the quark--photon fragmentation function
\cite{Glover:1993xc,ALEPH:1995zdi}. To achieve this, the definition of the
QED CS dipoles at this order is modified to include a fragmentation component
and an explicit dependence on the $z^{\rm cut}_\gamma$ cut, as described in \citeres{Denner:2014ina,Denner:2010ia,Denner:2009gj}.

\begin{figure}
        \centering
        \subfigure[\label{fig:as1a6_photon_loop}]{\includegraphics[width=0.49\linewidth]{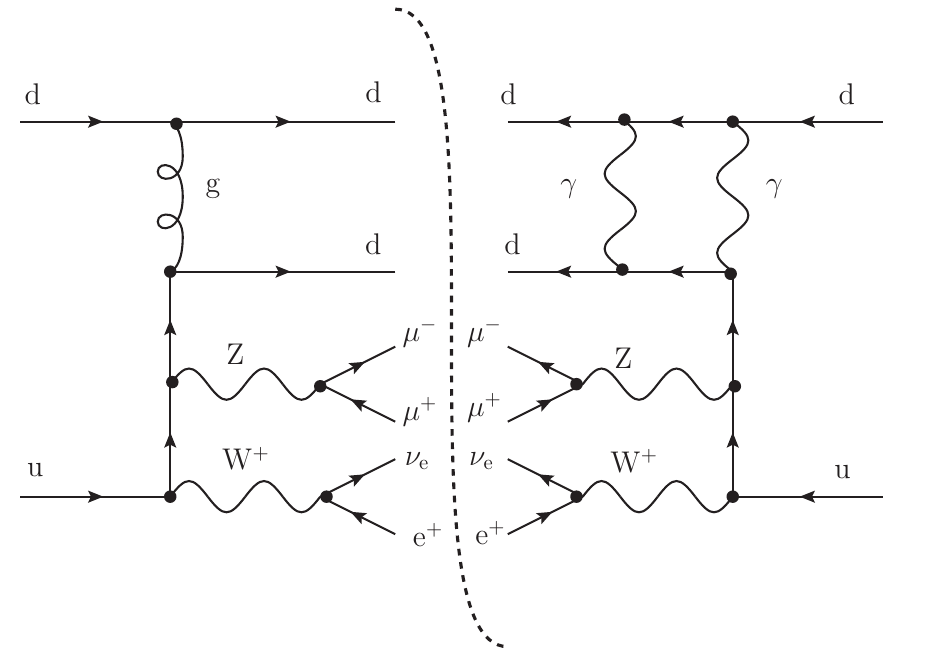}}
        \subfigure[\label{fig:as1a6_mixed_loop}]{\includegraphics[width=0.49\linewidth]{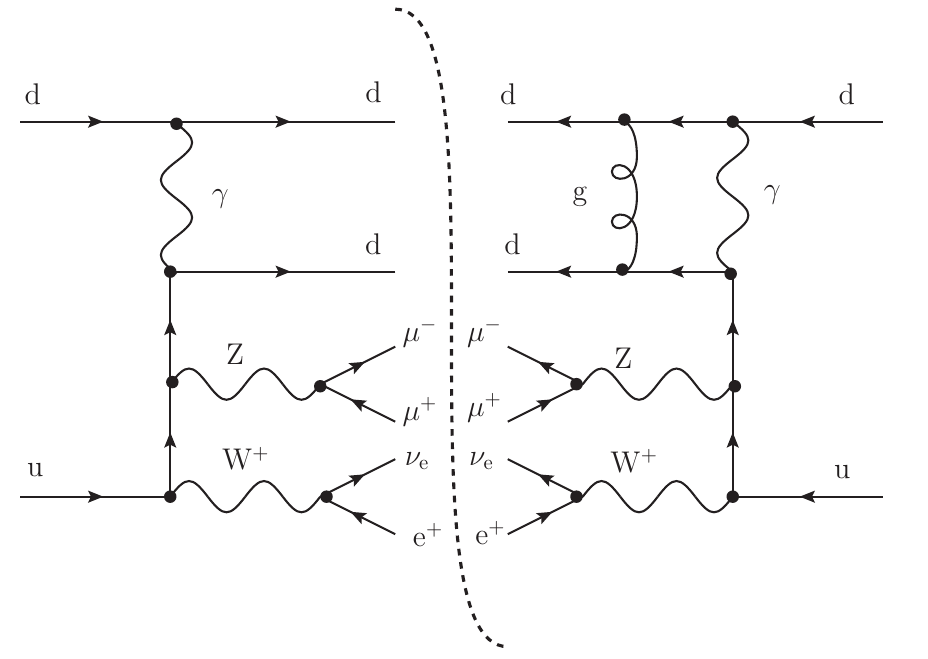}}
        \caption{Exemplary one-loop contributions of $\mathcal{O}(\alphas\alpha^6)$. While Fig.~\ref{fig:as1a6_photon_loop} unambiguously corresponds to an EW correction 
        to the $\order{\alphas\alpha^5}$, the contribution depicted in \reffi{fig:as1a6_mixed_loop} can be interpreted either as a QCD correction to the $\mathcal{O}(\alpha^6)$
        or an EW  correction to the $\mathcal{O}(\alphas\alpha^5)$.}\label{fig:as1a6_virt}
\end{figure}

The distinction between NLO QCD corrections to $\mathcal{O}(\alpha^6)$ and NLO EW corrections to
$\mathcal{O}(\alphas\alpha^5)$ that we made for the real contributions 
is not possible anymore for virtual ones, as has been discussed for
instance in~\citere{Biedermann:2017bss}.
Interferences of $\mathcal{O}(g^8)$ one-loop amplitudes with $\mathcal{O}(\gs^2 g^4)$ tree-level ones
[shown in \reffi{fig:as1a6_photon_loop}]
can be uniquely identified as virtual EW corrections to $\mathcal{O}(\alphas\alpha^5)$. Conversely,
interferences of $\mathcal{O}(g^6)$ tree level amplitudes with $\mathcal{O}(\gs^2 g^6)$ one-loop ones
can be considered either as QCD corrections to $\mathcal{O}(\alpha^6)$ or as EW corrections to
$\mathcal{O}(\alphas\alpha^5)$, owing to the presence of a QCD--EW mixed loop, as exemplified in
\reffi{fig:as1a6_mixed_loop}.

%%% Local Variables: 
%%% mode: latex
%%% TeX-master: "wvz_paper"
%%% End: 

%% file: results.tex
\section{Numerical results}\label{sec:numresults}
\subsection{Input parameters}\label{sec:input}
In the following, we present results for the LHC at a centre-of-mass energy of 13.6\TeV.  We consider
the process in Eq.~\refeqf{eq:procdef} with  two different-flavour charged leptons in the final
state and all leptons assumed to be massless.
%% We note that our results can be used as well 
%% for the case of identical leptons in the final state after applying
%% appropriate symmetry factor $1/2$ up to interference effects. These
%% interference effects are not doubly resonant and suppressed by factors
%% $\Gt/\Mt$. 
We work in the five-flavour scheme and therefore treat all light and bottom quarks as massless.
A unit quark-mixing matrix is understood.

The on-shell values for the masses and widths of the EW gauge bosons
are chosen according to \citere{Workman:2022ynf},
\begin{alignat}{2}
        \Mwo &= 80.377  \GeV \,, \qquad \;\, \Gwo&&=2.085  \GeV \,, \nnb\\
        \Mzo &= 91.1876 \GeV \,, \qquad \Gzo&&= 2.4952 \GeV \,,
\end{alignat}
and then converted to their pole values by means of the relations \cite{Bardin:1988xt}
\beq
\Mv = \frac{\Mvo}{\sqrt{1+({\Gvo}/{\Mvo})^2}}\,,\qquad\, 
\Gv = \frac{\Gvo}{\sqrt{1+({\Gvo}/{\Mvo})^2}}\,.
\eeq
The top-quark and Higgs-boson masses and widths are fixed as \cite{Workman:2022ynf}
\begin{alignat}{2}
        \label{eq:masswidthtophiggs}
        \Mt  &= 172.69 \GeV \,, \qquad \Gt\, &&= 1.42   \GeV \,, \nnb\\
        \MH  &= 125.25 \GeV \,, \qquad \GH\, &&= 0.0041 \GeV \,.
\end{alignat}
%
%% The top-quark width at LO has been computed with the
%% formulas of \citere{Jezabek:1988iv} and using the pole mass and width
%% for the $\PW$~boson as input. In order to meet the perturbative accuracy addressed
%% in this work, the NLO width has been
%% obtained upon applying QCD- and EW-correction factors from~\citere{Basso:2015gca}
%% to the LO width. All LO and NLO results in Sections~\ref{sec:integrated} and~\ref{sec:differential}
%% are obtained by using the NLO top-quark width in \refeq{eq:masswidthtop}.
The EW coupling is extracted from the Fermi constant $G_\mu$ by
means of \cite{Denner:2000bj,Dittmaier:2001ay}
\beq
\alpha = \frac{\sqrt{2}}{\pi}\,G_\mu\,\biggl|\mu_\PW^2\biggl(1-\frac{\mu_\PW^2}{\mu_\PZ^2}\biggr)\biggr|\,,
\eeq
where $G_\mu = 1.16638\cdot10^{-5} \GeV^{-2}$. Complex masses $\mu_V^2=M_{\rm V}-\ri\Gamma_{\rm V}M_{\rm V}$
enter the $\alpha$~definition consistently with the complex-mass scheme
\cite{Denner:1999gp,Denner:2005fg,Denner:2006ic,Denner:2019vbn},
where the masses of unstable particles, \ie the EW vector bosons
and the top quark, are treated as complex parameters in all
parts of the computation. As a consequence, the EW mixing
angle and the related couplings are complex valued as well.

For both the LO and the NLO calculation, we use \texttt{NNPDF40\_nnlo\_as\_01180\_qed}\, 
PDFs \cite{NNPDF:2024djq},
extracted at NNLO with $\as(\Mz)=0.118$.  The usage of this PDF set allows us
to properly account for the photon PDF. 
The strong coupling constant
$\as$ used in the calculation of the amplitudes matches the one used
in the evolution of PDFs.  The PDFs and the running of $\as$
are obtained by interfacing \mocanlo with {\scshape LHAPDF6}
\cite{Buckley:2014ana}. 

The QCD partons with pseudorapidity $|\eta|<5$ are clustered into jets by means of
the anti-$k_{\rm T}$ algorithm \cite{Cacciari:2008gp} with resolution
radius $R=0.4$. Photons are never considered as jets within our reconstruction.
We assume a perfect $\Pb$-jet tagging veto, which allows us to neglect
all contributions with bottom quarks in the final state.

Our fiducial region is defined by a choice of selection cuts which is inspired
by the HL LHC prospect studies for the ATLAS detector \cite{ATLAS:2018iou}.  
In particular, we require that %among the set of jets only two of them pass the cuts
exactly two jets pass the tagging cuts
\beq\label{eq:setup1}
\pt{\Pj_i} > 40 \GeV\,\quad\text{and}\quad |y_{\Pj_i}|<3\quad\text{with $i=1,\,2$} \,.
\eeq
We remark that these requirements on the jets effectively veto any additional jet, 
except if it has very high rapidity or very low transverse momentum. 

The two tagged jets $\Pj_1$ and $\Pj_2$ are required to fulfil 
\beq\label{eq:setup2}
50\GeV<M_{\Pj_1\Pj_2} < 100  \GeV\,.
\eeq

Photons are recombined with quarks and charged leptons using the anti-$k_{\rm T}$ clustering algorithm using $R=0.1$.
For the three charged leptons, we impose the rapidity cut
\beq\label{eq:setup3}
|y_{\Pl_i} |<4\,,
\eeq
where $\Pl_i$ runs over the set of charged leptons ordered by transverse momentum. 
The charged leptons are additionally required to satisfy
\beq\label{eq:setup4}
\pt{\Pl_1}\,>\,50\GeV\,,\qquad  \pt{\Pl_2}\,>\,40\GeV\,, \qquad \pt{\Pl_3}\,>\,20\GeV\,.
\eeq
The invariant mass of the pair of opposite-sign same-flavour leptons,
\ie the $\mu^+\,\mu^-$ pair in our case, must fall into the mass window
\beq\label{eq:setup5}
76\GeV< M_{\mu^+\mu^-} <106\GeV\,,
\eeq
while the positron $\Pe^+$ is constrained by the transverse-mass cut
\beq\label{eq:setup6}
\tm{\PW^+}>20\GeV\, \quad\text{with}\quad \tm{\PW^+}=\sqrt{2\,\pt{\Pe^+}\pt{\nu_\text{e}}\,(1-\cos(\phi_{\Pe^+}-\phi_{\nu_\text{e}}))}\,,
\eeq
where $\phi$ is the azimuthal angle around the collision axis. 
We note that the definition of $\tm{\PW^+}$ makes use of the momentum of the neutrino $\nu_\text{e}$ at Monte Carlo-truth level, 
which we employ as a definition for the missing transverse momentum.  
%, and remark that our selection cuts imply the requirement of a partonic center-of-mass  energy of at least $190\GeV$.

We set  the factorisation and renormalisation scales to the same central value $\mu_0$,
%\ie $\mu_{\rm R}=\mu_{\rm F}=\mu_0$. %For $\mu_0$ we made use of the following dynamical choice:
%For $\mu_0$ we choose the fixed value:
\beq\label{eq:scaleA}
\mu_{\rm F}=\mu_{\rm R}=\mu_0 = \Mzo \,.
\eeq

%The uncertainties in our results are estimated by computing the $7$-point scale envelope, \ie
%by considering the maximum and minimum values of the cross~section evaluated on the set
%of scales $(\mu_{\rm R},\,\mu_{\rm F})$ defined as
%\beq\label{eq:scaleset}
%\biggl(\frac{\mu_{\rm R}}{\mu_0},\,\frac{\mu_{\rm F}}{\mu_0}\biggr)\in\{(0.5,0.5)\,,(0.5,1)\,,(1,0.5)\,,(1,1)\,,(2,1)\,,(1,2)\,,(2,2)\}\,.
%\eeq
%Note that the $\as$ coupling entering the calculation of the NLO top-quark width is kept fixed
%for the evaluation of the scale envelope.
%
%In the following two sections, we present results for the fiducial cross~sections and
%differential distributions where all contributions described in \refse{sec:calcdetails}
%are combined in an \emph{additive} scheme, \ie
%\begin{align}
%       \label{eq:addscheme}
%       \sigma_{\rm LO+NLO} ={}& \sigma_{\rm LO_1}\,+\,\sigma_{\rm NLO_1} 
%       \,+\,\sigma_{\rm LO_2}\,+\,\sigma_{\rm NLO_2}\,,
%       %    \,+\,\sigma_{\rm LO_3}\,+\,\sigma_{\rm NLO_3}+\,\sigma_{\rm NLO_4}\,,
%\end{align}
%which provides an exact result at the order of truncation of the perturbative expansion.

\subsection{Fiducial cross sections}\label{sec:integrated}

To assess the signal-to-background ratio, we compare the cross sections of the different leading orders discussed in \refse{sec:calcdetails},
integrated over the phase space defined by the cuts of \refse{sec:input}. Table \ref{tab:LO_ordercomp_all} shows these results, where we 
include all partonic processes that arise at the given order except those with one or more bottom quarks in the final state. 
\begin{table}%[tbp]
        \centering
        \pgfplotstableset{% global config, for example in the preamble
                % these columns/<colname>/.style={<options>} things define a style
                % which applies to <colname> only.
                %               columns/type/.style={string type,column type=c,column name=\textsc{Subprocess type}},
                %columns/type/.style={string type,column type=c,column name=\textsc{Subprocess}},
                columns/type/.style={string type,column type=c,column name=},
                columns/as0a6/.style={
                        string type,
                        column name=$\mathcal{O}(\alpha^6)$,
                        string replace={0.0}{}
                },
                columns/as1a5/.style={
                        string type,
                        column name=$\mathcal{O}(\alphas\alpha^5)$,
                        string replace={0.0}{}
                },
                columns/as2a4/.style={
                        string type,
                        column name=$\mathcal{O}(\alphas^2\alpha^4)$,
                        string replace={0.0}{}
                },
                columns/xs_sum/.style={string type,column type=c,column name=sum},
                empty cells with={--}, % replace empty cells with '--'
                every head row/.style={before row=\toprule,after row=\midrule},
                %               every last row/.style={before row=\midrule,after row=\bottomrule}
                every last row/.style={after row=\bottomrule}
        }
        \begin{filecontents*}{tables/LO_ordercomp_all.csv}
type,as0a6,as1a5,as2a4,xs_sum
$\sigma_{\rm LO} \;[{\rm ab}]$,$50.230(2)$,$8.144(2)$,$847.7(5)$,$906.0(5)$
$\Delta_{\rm tot} \;[\%]$,$5.54$,$0.90$,$93.56$,$100.00$
\end{filecontents*}
\pgfplotstabletypeset[ % local config, applies only for this table
        col sep=comma,
        1000 sep={\,},
        columns/info/.style={
                fixed,fixed zerofill,precision=1,showpos,
                column type=r,
        }
        ]
        {tables/LO_ordercomp_all.csv}
        \caption{
                Cross sections (in ab) of $\Pp\Pp\to\mu^+\mu^-\Pe^+\nu_\Pe\,\Pj\,\Pj$ at each of the leading orders. 
                Integration errors on the last digit are given in parentheses. 
                The cross sections include all partonic subprocesses 
                that arise at the given order except those with one or more bottom quarks in the final state.
                The second row shows each LO cross section relative to
                the sum of the leading orders in percent.
        }
        \label{tab:LO_ordercomp_all}
\end{table}
The second entry of the middle row contains the LO cross section of our signal, which contributes to $\order{\alpha^6}$,
\begin{align}\label{eq:LOsignal}
        \sigma^{\rm sig}_{\rm LO} = 50.230(2) \;{\rm ab} \,,
\end{align}
where the integration uncertainty on the last digit is given in parentheses. 
In the bottom row, the relative size of each order is displayed in percent. While the interference background of $\order{\alphas\alpha^5}$ in the third column 
is below the percent level, it is evident that the QCD background, in
the fourth column, can not be reduced by our cut setup to less than 93.6\%.

We further study the signal-to-background ratio within each subprocess type. 
Table~\ref{tab:LO_ordercomp_all_merged} displays the contribution at each LO of each subprocess type to our signal. 
\begin{table}%[tbp]
        \centering
        \pgfplotstableset{% global config, for example in the preamble
                % these columns/<colname>/.style={<options>} things define a style
                % which applies to <colname> only.
                %               columns/type/.style={string type,column type=c,column name=\textsc{Subprocess type}},
                columns/type/.style={string type,column type=c,column name=Subprocess},
                columns/as0a6/.style={
                        string type,
                        column name=$\mathcal{O}(\alpha^6)$,
                        string replace={$0.0$}{}
                },
                columns/as1a5/.style={
                        string type,
                        column name=$\mathcal{O}(\alphas\alpha^5)$,
                        string replace={$0.0$}{}
                },
                columns/as2a4/.style={
                        string type,
                        column name=$\mathcal{O}(\alphas^2\alpha^4)$,
                        string replace={$0.0$}{}
                },
                columns/xs_sum/.style={string type,column type=c,column name=sum},
                empty cells with={--}, % replace empty cells with '--'
                every head row/.style={before row=\toprule,after row=\midrule},
                %               every last row/.style={before row=\midrule,after row=\bottomrule}
                every last row/.style={after row=\bottomrule},
                every odd row/.style={after row=\midrule}
        }
        \begin{filecontents*}{tables/LO_ordercomp_all_merged.csv}
type,as0a6,as1a5,as2a4,xs_sum
$\sigma_{\rm LO}^{qq/\bar{q}\bar{q}/q\bar{q}} \;[{\rm ab}]$,$48.657(2)$,$-2.4189(5)$,$59.2(1)$,$105.4(1)$
$\Delta_{\alpha^6} \;[\%]$,$96.87$,$-4.82$,$117.92$,$209.98$
$\sigma_{\rm LO}^\gamma \;[{\rm ab}]$,$0.8592(1)$,$10.563(2)$,--,$11.422(2)$
$\Delta_{\alpha^6} \;[\%]$,$1.71$,$21.03$,$0.0$,$22.74$
$\sigma_{\rm LO}^{{\rm b}\bar{{\rm b}}} \;[{\rm ab}]$,$0.7137(1)$,--,$0.0034516(7)$,$0.7171(1)$
$\Delta_{\alpha^6} \;[\%]$,$1.42$,$0.0$,$0.01$,$1.43$
$\sigma_{\rm LO}^{\rm b/\bar{b}} \;[{\rm ab}]$,$12.3879(6)$,--,$7.628(2)$,$20.016(2)$
$\Delta_{\alpha^6} \;[\%]$,$24.66$,$0.0$,$15.19$,$39.85$
\end{filecontents*}
\pgfplotstabletypeset[ % local config, applies only for this table
        col sep=comma,
        1000 sep={\,},
        columns/info/.style={
                fixed,fixed zerofill,precision=1,showpos,
                column type=r,
        }
        ]
        {tables/LO_ordercomp_all_merged.csv}
        \caption{
                Cross sections (in ab) at each of the leading orders by type of partonic subprocess of $\mu^+\mu^-\Pe^+\nu_\Pe\,\Pj\,\Pj$ production. 
                Here, $q$ stands for a generic light (anti)quark in $S_q$ [see Eq.~\eqref{eq:lo1_qqb}]. 
                The rightmost column shows the cross~section of each subprocess type summed over the leading orders. 
                The cross section of subprocesses with at least one
                photon in the initial state is denoted by
                $\sigma^\gamma_{\rm LO}$, while 
                $\sigma^{\rm b\bar{\Pb}}_{\rm LO}$ stands for the cross section of bottom--antibottom-initiated subprocesses. 
                For comparison, we show the cross section $\sigma^{\rm b}_{\rm LO}$ of subprocesses with at least one bottom-quark in the final state, 
                which is not part of our signal. Integration errors on the last digits are given in parentheses.
                The odd rows ($\Delta_{\alpha^6}$) show each cross section relative to $\sigma^{\rm sig}_{\rm LO}$ in Eq.~(\ref{eq:LOsignal}).} 
        \label{tab:LO_ordercomp_all_merged}
\end{table}
For comparison, it also shows the contributions $\sigma^\Pb_{\rm LO}$ of subprocesses with one or more bottom quarks in the final state, which are not part of our signal. 
The contribution $\sigma^\gamma_{\rm LO}$ stands for either double- or single-photon-induced subprocesses, depending on the perturbative order.  
Below each absolute number (in ab), we show the relative size $\Delta_{\alpha^6}$ of each contribution with respect to the total cross section of the signal in Eq.~(\ref{eq:LOsignal}).

We observe that quark--(anti)quark-initiated subprocesses make up $97\%$ of the cross section of our LO signal at $\order{\alpha^6}$. 
At $\order{\alphas\alpha^5}$, these subprocesses give rise to
interference contributions that are negative when integrated over the
fiducial phase-space volume
and represent less than 5\% relative to our signal. On the other hand, the $\order{\alphas^2\alpha^4}$ contribution of $qq/\bar{q}\bar{q}/q\bar{q}$-induced subprocesses 
is larger than the one of $\order{\alpha^6}$ by 21\%.

Photon-induced contributions arise only at orders $\order{\alpha^6}$
and $\order{\alphas\alpha^5}$. At the former order, only
photon--photon-initiated subprocesses are present and represent $2\%$
of the signal. At the latter order, only single-photon-initiated subprocesses exist.  
Contrary to the quark--(anti)quark subprocesses, their contribution arises from squared amplitudes and is as large as 21\% of our signal.

Bottom--antibottom-induced subprocesses contribute at orders $\order{\alpha^6}$ and $\order{\alphas^2\alpha^4}$, as discussed in \refse{sec:calcdetails}. 
At our LO signal, the $\Pb\bar{\Pb}$ cross section is similar in size to that of the $\gamma\gamma$-induced subprocesses. 
In both cases the PDF suppression is compensated by the
enhancement from EW resonances (see \reffi{fig:a6_aa_b}). That does not occur at $\order{\alphas^2\alpha^4}$, 
where the $\Pb\bar{\Pb}$ cross section is two orders of magnitude smaller.

Subprocesses with one or more bottom quarks in the final state, which we assume to be perfectly vetoed, would induce a sizeable contribution 
of $25\%$ of our signal at $\order{\alpha^6}$. At $\order{\alphas^2\alpha^4}$, their contribution is smaller and represents $15\%$ 
of our signal cross section. This can be explained by the enhancement from top-quark resonances, which are present at $\order{\alpha^6}$
but absent at $\order{\alphas^2\alpha^4}$, as discussed in \refse{sec:calcdetails}. As the $\Pb\bar{\Pb}$-induced subprocesses, 
those with one or more $\Pb$~quarks in the final state do not contribute to $\order{\alphas\alpha^5}$ because of colour algebra.

Table~\ref{tab:LO2} displays a more detailed decomposition of each LO, where the contributions 
of all different subprocess types are shown separately. 
\begin{table}[tbp]
        \centering
        \pgfplotstableset{% global config, for example in the preamble
                % these columns/<colname>/.style={<options>} things define a style
                % which applies to <colname> only.
                %               columns/type/.style={string type,column type=c,column name=\textsc{Subprocess type}},
                columns/type/.style={string type,column type=c,column name=Subprocess},
                columns/as0a6/.style={
                        string type,
                        column name=$\mathcal{O}(\alpha^6)\;[\rm{ab}]$,
                },
                columns/as1a5/.style={
                        string type,
                        column name=$\mathcal{O}(\alphas\alpha^5)\;[\rm{ab}]$,
                },
                columns/as2a4/.style={
                        string type,
                        column name=$\mathcal{O}(\alphas^2\alpha^4)\;[\rm{ab}]$,
                },
                empty cells with={--}, % replace empty cells with '--'
                every head row/.style={before row=\toprule,after row=\midrule},
                every last row/.style={before row=\midrule,after row=\bottomrule}
        }
        \begin{filecontents*}{tables/LO.csv}
type,as0a6,as1a5,as2a4
$q\bar{q}\rightarrow q\bar{q}$,$47.616(2)$,$-0.4524(2)$,$39.1(1)$
${\rm b}$,$12.3879(6)$,--,$7.628(2)$
$qq/\bar{q}\bar{q}$,$1.04105(5)$,$-1.9664(4)$,$20.05(7)$
$\gamma\gamma$,$0.8592(1)$,--,--
${\rm b}\bar{{\rm b}}$,$0.7137(1)$,--,$0.0034516(7)$
$\gamma q/\gamma\bar{q}$,--,$9.617(2)$,--
$\Pg\gamma$,--,$0.9460(4)$,--
$\Pg\Pg$,--,--,$17.5(1)$
$\Pg q/\Pg\bar{q}$,--,--,$608.8(3)$
$q\bar{q}\rightarrow \Pg\Pg$,--,--,$162.0(1)$
total,$62.618(2)$,$8.144(2)$,$855.3(5)$
\end{filecontents*}
\pgfplotstabletypeset[ % local config, applies only for this table
        col sep=comma,
        1000 sep={\,},
        columns/info/.style={
                fixed,fixed zerofill,precision=1,showpos,
                column type=r,
        }
        ]
        {tables/LO.csv}
        \caption{
                Cross sections (in ab) of all sets of partonic subprocesses contributing 
                to each of the leading orders of $\mu^+\mu^-\Pe^+\nu_\Pe\,\Pj\,\Pj$ production. 
                The label $q(\bar{q})$ stands for a generic light (anti)quark in $S_q$.
                In the last row, the sum of all partonic channels contributing at that specific order
                is reported. Integration errors on the last digit are given in parentheses.
        }
        \label{tab:LO2}
\end{table}
To simplify this presentation, we include all subprocesses induced by one quark and one antiquark in the type $q\bar{q}$ in the second row. 
This type thus contains both subprocesses that are compatible with a tri-boson topology, like those in Eq.~(\ref{eq:lo1_qqb}), 
as well as subprocesses that are not, as those on the right of Eq.~(\ref{eq:lo1_qq}). 
The subprocess type $qq/\bar{q}\bar{q}$ corresponds to the first two
partonic process types of \refeq{eq:lo1_qq}, which are part of the tri-boson background.
They are listed separately in the fourth row of Table~\ref{tab:LO2}. 

The second column of \refta{tab:LO2} confirms once more that 
the quark--antiquark subprocesses are responsible for most of our signal cross section. This is a 
consequence of our fiducial phase space, as this class of subprocesses genuinely embeds the tri-boson 
signature (see \refse{sec:calcdetails}) and confirms that our selection cuts work as intended.

In the third column, the interference contributions of $qq$-, $\bar{q}\bar{q}$-, and $q\bar{q}$-induced subprocesses are 
both negative and of the same order of magnitude. In contrast, there is a difference between the two types of single-photon-induced 
subprocesses, $\gamma q/\gamma\bar{q}$ and $\Pg \gamma$, the former contribution being an order of magnitude larger than the latter. 
The likely reason for this is our event selection, as has been discussed in \refse{sec:lo2}. 
The $\gamma q/\gamma\bar{q}$ initial state produces a (anti)quark--gluon
pair, which is more likely to be misidentified as the decay products
of a vector boson than a quark--antiquark pair that is 
produced by $\Pg \gamma$-induced subprocesses. %\footnote{Are we sure?}.
This is because the invariant mass of the quark--gluon pair is typically smaller than the one of the quark--antiquark pair in these subprocesses.

The rightmost column of Table~\ref{tab:LO2} shows that the by far largest contribution to the $\order{\alphas^2\alpha^4}$ cross section 
stems from gluon--quark-induced subprocesses. It represents more than
70\% of this order, followed by the quark--antiquark-induced
contribution with a final-state gluon pair, which makes up 19\% of the
full $\order{\alphas^2\alpha^4}$ result.  
As before, we attribute these large contributions to the misidentification of a $\Pg q/\Pg\bar{q}$ or $\Pg\Pg$ pair as a 
hadronically decaying vector boson. Gluon--gluon-induced subprocesses, which are expected to be enhanced by their PDFs, 
produce a final-state quark pair that is misidentified less often owing to its large invariant mass (see \refse{sec:lo3}), which likely leads to the relatively small contribution of $17.5\ab$. 
All other subprocesses that contribute at this order have no external gluons but proceed by the exchange of an internal gluon 
as mentioned in \refse{sec:calcdetails}. The corresponding lack of any
enhancement mechanism for these latter channels offers a possible explanation for the difference in size between the
$q\bar{q}$-induced subprocesses with $\Pg\Pg$ and $q\bar{q}$ final states.

Table~\ref{tab:NLO_perc} displays all the EW and NLO QCD corrections that we calculated. 
As before, these results are decomposed by type of subprocess. 
\begin{table}[tbp]
        \centering
        \pgfplotstableset{% global config, for example in the preamble
                % these columns/<colname>/.style={<options>} things define a style
                % which applies to <colname> only.
                %               columns/type/.style={string type,column type=c,column name=\textsc{Subprocess type}},
                columns/type/.style={string type,column type=c,column name=Subprocess},
                columns/as0a6/.style={
                        string type,
                        column name=$\mathcal{O}(\alpha^6)\;[\rm{ab}]$,
                },
                columns/as0a7/.style={
                        string type,
                        column name=$\mathcal{O}(\alpha^7)\;[\rm{ab}]$,
                        string replace={0.0}{},
                },
                columns/as0a7_ov_as0a6/.style={
                        string type,
                        %                       column name=$\mathcal{O}(\alpha^7)/\mathcal{O}(\alpha^6)\;[\%]$,
                        column name=$\Delta_{\alpha^6}^{(i)}\;[\%]$,
                        string replace={$0.0$}{},
                },
                columns/as1a6/.style={
                        string type,
                        column name=$\mathcal{O}(\alphas\alpha^6)\;[\rm{ab}]$,
                        string replace={0.0}{},
                },
                columns/as1a6_ov_as0a6/.style={
                        string type,
                        %                       column name=$\mathcal{O}(\alphas\alpha^6)/\mathcal{O}(\alpha^6)\;[\%]$,
                        column name=$\Delta_{\alpha^6}^{(i)}\;[\%]$,
                        string replace={$0.0$}{},
                        string replace={$0.0$}{},
                },
                empty cells with={--}, % replace empty cells with '--'
                every head row/.style={before row=\toprule,after row=\midrule},
                every last row/.style={before row=\midrule,after row=\bottomrule}
        }
        \begin{filecontents*}{tables/NLO_perc.csv}
type,as0a6,as0a7,as0a7_ov_as0a6,as1a6,as1a6_ov_as0a6
$q\bar{q}\rightarrow q\bar{q}$,$47.616(2)$,$-7.81(5)$,$-16.4$,$-6.83(6)$,$-14.3$
$qq/\bar{q}\bar{q}$,$1.04105(5)$,$-0.1156(3)$,$-11.1$,$4.19(1)$,$402.5$
$\gamma\gamma$,$0.8592(1)$,$-0.1045(6)$,$-12.1$,$-0.1341(4)$,$-15.6$
${\rm b}\bar{{\rm b}}$,$0.7137(1)$,$-0.0836(4)$,$-11.7$,$-0.2195(5)$,$-30.7$
$\gamma q/\gamma\bar{q}$,--,$0.9175(2)$,$0.0$,$-0.593(6)$,$0.0$
$\Pg\gamma$,--,--,$0.0$,$0.0053(7)$,$0.0$
$\Pg q/\Pg\bar{q}$,--,--,$0.0$,$5.766(3)$,$0.0$
total,$50.230(2)$,$-7.20(5)$,$-14.3$,$2.17(6)$,$4.3$
\end{filecontents*}
\pgfplotstabletypeset[ % local config, applies only for this table
        col sep=comma,
        1000 sep={\.},
        columns/info/.style={
                fixed,fixed zerofill,precision=1,showpos,
                column type=r,
        }
        ]
        {tables/NLO_perc.csv}
        \caption{
                Total and relative NLO cross~sections (in ab) for different types of partonic subprocesses contributing to
                $\mu^+\mu^-\Pe^+\nu_\Pe\Pj\Pj$ production at the LHC. 
                The second column shows the LO signal from Table~\ref{tab:LO2} for reference. The third and fifth columns list
                the NLO EW and QCD corrections, respectively. The fourth and sixth columns show the NLO corrections relative to 
                the LO signal of each subprocess type from the second column.   
                In the last line, the sum of all partonic channels contributing at that specific order
                is reported. Integration errors on the last digit are given in parentheses.
        }
        \label{tab:NLO_perc}
\end{table}
The second column reproduces the LO signal results from Table~\ref{tab:LO2} for comparison. 
To the right of each absolute NLO correction, we give its size $\Delta_{\alpha^6}^{(i)}$ 
relative to the $\order{\alpha^6}$ cross section of the same subprocess type. 

The $q\bar{q}$-induced subprocesses receive the largest EW contributions, which amount to $-16.4\%$ 
of the corresponding leading order and $-15.5\%$ of the total $\order{\alpha^6}$ cross section.
The rest of the subprocess types gets relative EW corrections of $-12\%$, which are nevertheless below 
the percent level with respect to the total LO signal cross section. 
The $\order{\alpha^7}$ cross section of photon--quark-induced subprocesses, 
for which no contribution at $\order{\alpha^6}$ exists, amounts to about $+1.8\%$ of our full LO signal. 
All in all, our process receives relative EW corrections of $-14.3\%$,
which is almost as large as for the same final state in the VBS
phase space \cite{Denner:2019tmn}.

In previous studies, it has been found for several tri-boson production processes that the quark-induced and 
photon-induced NLO EW contributions of order $\order{\alpha^7}$ can be individually large but often cancel 
against each other to yield more modest total NLO EW corrections. 
In \citere{Nhung:2013jta}, where the on-shell production of $\PW^+\PW^-\PZ$ at the LHC was studied,
a relative EW correction of $-8.8\%$ was reported for the quark-induced channels and $+6.8\%$ for 
the photon--quark-induced ones. 
%In~\cite{Schonherr:2018jva}, predictions for off-shell tri-boson
%production with fully-leptonic decays were presented. There, Table 4 cites the relative EW corrections  
%of $-6.1\%$ and $+3.2\%$ for quark- and photon--quark-induced subprocesses, respectively, of 
%the final state $\Pe^- \Pe^+ \mu^+ \nu_\mu \nu_\mu \bar{\nu}_\mu$, which is compatible with $\PW\PZ\PZ$ 
%production. 
%
Relative EW corrections have been calculated for $\PW\PW\PW$ production with different final states. In \citere{Dittmaier:2017bnh}, 
the on-shell process was investigated and the values $-4.17\%$ and $+11.46\%$ for quark- and photon--quark-induced contributions were found. 
The off-shell process with fully leptonic final state was studied in \citere{Dittmaier:2019twg} for two different cut setups.
There, for the relative EW corrections from quark- and photon--quark-induced contributions, respectively, 
the values $-8.5\%$ and $+2.4\%$ were obtained for one setup (Table 2), while the values $-7.8\%$ and $+7.78\%$ were found for the other setup (Table 4).
The semi-leptonic final state of $\PW\PW\PW$ production was investigated in \citere{Denner:2024ufg}, where 
for the quark- and photon--quark-induced corrections the values $-7.2\%$ and $+2.6\%$ were found.

The extent of the cancellation between quark- and photon--quark-induced contributions at $\order{\alpha^7}$ 
appears to be cut dependent. For the process studied in this article, it is not substantial, as mentioned above.
On the other hand, the relative contribution of quark-induced processes at $\order{\alpha^7}$ that we find 
exceeds the values in the literature for related tri-boson processes. 
Large negative EW corrections at high energies typically arise from large EW Sudakov logarithms. 
In our setup, the mean partonic centre-of-mass energy $\langle\sqrt{\hat{s}}\rangle$ ranges between $700$ and $770\GeV$ for some of the most 
significant partonic channels. In contrast, this quantity turns out to be between $440$ and $600\GeV$ in 
$\PW\PW\PW$ production with a semi-leptonic final state using the setup of \citere{Denner:2024ufg}. 
The comparatively high value of $\langle\sqrt{\hat{s}}\rangle$ in our
setup is at least partly due 
to our selection cuts, as we have confirmed by varying them. 
While no single cut is responsible for the large $\langle \sqrt{\hat{s}}\rangle$, 
the relatively high lower bounds on the transverse momenta and
invariant masses enhance the average partonic centre-of-mass energy. 
In contrast to the typically large corrections of $-15\%$ for VBS
processes
\cite{Biedermann:2016yds,Denner:2019tmn,Denner:2020zit,Denner:2022pwc},
the corrections of $-14\%$ for $\PW^+\PZ\Pj\Pj$
production considered here cannot be reproduced by a simple Sudakov
approximation neglecting angular-dependent logarithms. Such an
approximation would in fact result in even larger EW corrections. Obviously, the
Sudakov limit, where all invariants are large compared to $\MW^2$, is not
dominant in the relevant phase space for the considered process.

Turning back to \refta{tab:NLO_perc}, the $\order{\alphas\alpha^6}$ correction to the $q\bar{q}$-induced subprocesses is similar in size to 
the corresponding EW correction and also negative. For $qq/\bar{q}\bar{q}$-induced subprocesses, we find a very large, 
positive QCD correction of about $400\%$ with respect to the $\order{\alpha^6}$ cross section of this subprocess type. 
As mentioned in \refse{sec:calcdetails}, $qq/\bar{q}\bar{q}$-induced subprocesses are
not compatible with genuine tri-boson contributions 
and are efficiently suppressed in our setup at LO. 
The large $\order{\alphas\alpha^6}$ contribution is caused by bypassing our selection cuts, 
via the radiation of real gluons (see \refse{sec:nlo2}).
When an additional final-state gluon is present and well separated from the other two QCD partons [see \refeq{eq:nlo2_qq}], it can be misidentified 
as one of the two tagging jets coming from the hadronic decay of a
vector boson (see also \citere{Denner:2024ufg}). 
If the third QCD parton has a large rapidity and low enough transverse momentum, it is not tagged and the event 
is not dismissed by our requirement of exactly two tagged jets (see
also \refse{sec:nlo2}). 

Channels induced by two photons receive an $\order{\alphas\alpha^6}$ correction of $-16\%$ and bottom--antibottom-induced 
subprocesses an even larger one at $-31\%$. In both cases, the
$\order{\alphas\alpha^6}$ correction 
represents less than a percent of the full LO signal. We investigate the QCD corrections to the $\Pb\bar{\Pb}$-induced 
subprocesses in more detail because of their large relative size. 
We begin by comparing the subprocesses $\Pb\,\bar{\Pb}\to\mu^+\mu^-\Pe^+\nu_\Pe\,\Pc\,\bar{\Ps}$
and  \mbox{$\Pd\,\bar{\Pd}\to\mu^+\mu^-\Pe^+\nu_\Pe\,\Pc\,\bar{\Ps}$}, whose
matrix elements differ by the presence of top-quark 
propagators that arise only for the former subprocess. While the real and dipole contributions 
are similar for both subprocesses relative to the corresponding LO, the relative virtual contributions differ 
substantially. When artificially setting the 
top-quark mass to zero in the $\Pb\bar{\Pb}$-induced subprocesses, the relative real and dipole contributions remain almost unaffected, 
but the relative virtual contribution significantly changes. Thus, the 
large relative QCD corrections to the $\Pb\bar{\Pb}$-induced
subprocesses are due to top-mass effects in the virtual amplitudes.

Finally, the initial states $\gamma q/\gamma\bar{q}$, $\Pg \gamma$, and $\Pg q/\Pg\bar{q}$ appear as real-correction subprocesses at $\order{\alphas\alpha^6}$. 
The contribution of the latter is positive and amounts to $11.5\%$ of the total LO signal, while the former two are below the percent level.
The substantial cancellations that occur at this order yield an $\order{\alphas\alpha^6}$ correction of altogether $+4.3\%$. 

\subsection{Differential cross sections}\label{sec:differential}

%%%%%%%%%%%%%%%% Discussion of differential partonic COM energy
In this subsection, we present differential results for the process $\Pp \Pp \to \mu^+\mu^-\Pe^+\nu_\Pe\Pj\Pj$ at 
orders $\order{\alpha^6}$, $\order{\alpha^7}$, and $\order{\alphas\alpha^6}$. 
From now on, LO will refer to the contribution of $\order{\alpha^6}$, while NLO EW and NLO QCD denote 
contributions of $\order{\alpha^6} + \order{\alpha^7}$ and $\order{\alpha^6} + \order{\alphas\alpha^6}$, respectively. 
We call the combination $\order{\alpha^6} + \order{\alpha^7} + \order{\alphas\alpha^6}$ total NLO. 
We remark that NLO QCD quantities also contain EW corrections to the non-signal LO $\order{\alphas\alpha^5}$.  
Furthermore, only the subprocess types listed in Table~\ref{tab:NLO_perc} are taken into account and are labelled here with the same notation.
% and we employ that same notation for the subprocess types. 
 
\begin{figure}
        \centering
        \includegraphics[width=0.497\textwidth,page=1]{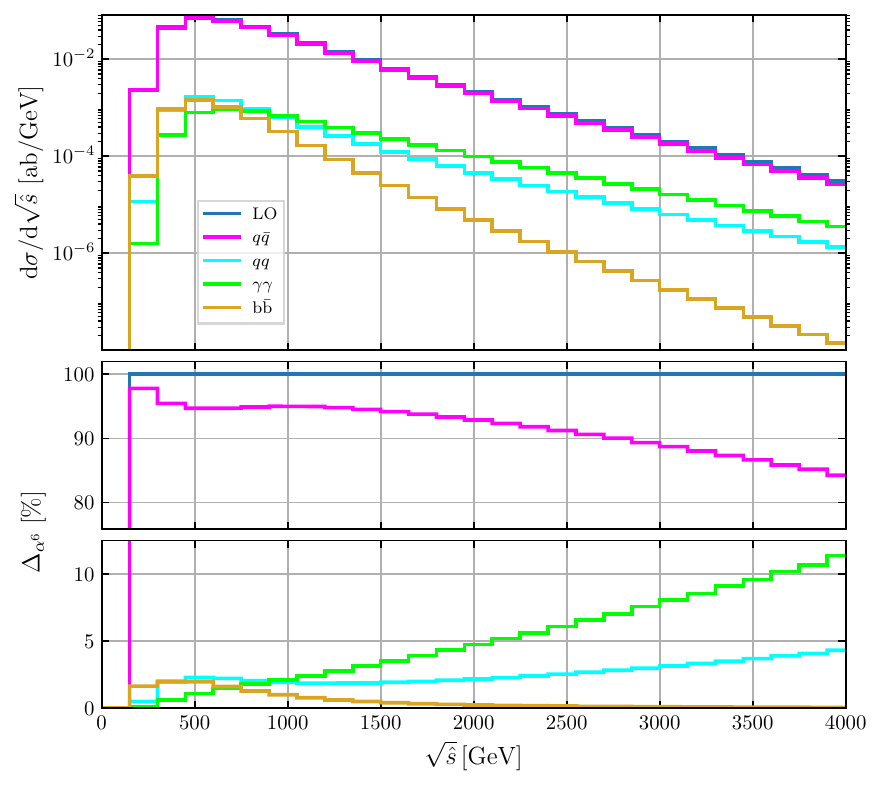}% cms energy
        \includegraphics[width=0.503\textwidth,page=1]{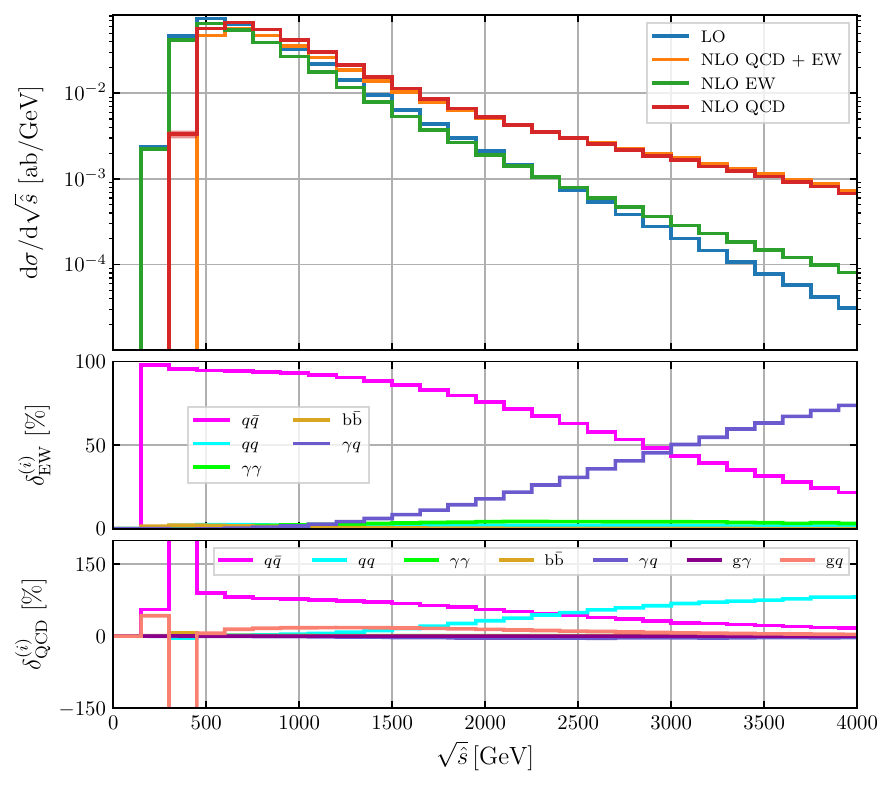}% cms energy
        \caption{Distributions of the partonic centre-of-mass energy,
          $\sqrt{\hat{s}}$, at LO $\order{\alpha^6}$ (left) and NLO
          $\order{\alpha^6}+\order{\alpha^7}$ and
          $\order{\alpha^6}+\order{\alphas\alpha^6}$ (right)
                decomposed into the contributions of different subprocess types. See text for details about the lower panels.
        }
        \label{fig:cms_bytype}
\end{figure}

We begin our discussion with  the differential cross sections with respect 
to the partonic centre-of-mass-energy $\sqrt{\hat{s}}$ in \reffi{fig:cms_bytype}. While this quantity is not experimentally accessible, its study offers
insights into the individual contributions of each subprocess type. 

On the left plot, the $\order{\alpha^6}$ cross section is shown in blue, decomposed in the contributions from 
different types of partonic subprocesses. The middle and bottom panels display the relative sizes of each contribution in percent, 
zoomed into separate ranges for visibility. 
On this plot, we recognise the dominance of the $q\bar{q}$-induced subprocesses that we observed at the integrated 
level in \refse{sec:integrated} but also find that it becomes less pronounced at higher values of the partonic centre-of-mass energy. 
The relative size of the $q\bar{q}$ contribution decreases from $98\%$ at low energies to $85\%$ at $4\TeV$. 
To make up for it, the $\gamma\gamma$- and $qq/\bar{q}\bar{q}$-induced relative contributions grow to more than 10\% and just below 5\%
in the same energy range, respectively. While the growth of the relative $\gamma\gamma$ contribution can be attributed to the 
photon PDF, we interpret the rise of the relative $qq/\bar{q}\bar{q}$ contribution as
a decrease in efficiency of our tri-boson selection cuts in favour of
VBS contributions.

We turn to the right plot of \reffi{fig:cms_bytype}. On its top panel, the NLO EW and QCD cross sections are shown separately 
(in green and red, respectively) and combined (in orange). 
We observe large negative QCD corrections in the first three bins up to $450\GeV$. While the LO curve is positive in the second bin, 
between $150$ and $300\GeV$, the NLO QCD curve is in fact negative
(not visible owing to the logarithmic scale). 
Furthermore, there is a gap between these curves of one order of magnitude in the bin between $300$ and $450\GeV$.
After the fourth bin, these corrections turn positive and continue
growing towards higher energies, where the NLO QCD distribution significantly exceeds the LO one. 

We attribute the large QCD corrections at low partonic energies,
$\langle\sqrt{\hat{s}}\rangle\lesssim 500\GeV$, to the phase-space suppression of the real contribution 
close to the on-shell threshold, given by $2\MW + \MZ \approx
250\GeV$ and the lack of cancellations between the real and virtual
corrections. As can be seen in the upper right panel of \reffi{fig:cms_bytype}, 
the cross section is dominated by the Born-like contributions below 500\GeV, which include integrated and subtraction dipoles and virtual corrections, 
and are mostly supported in the low-energy regime. The real corrections mainly contribute at higher energies, 
where there is more phase space available for the real emission. 
Furthermore, we find that $\Pg q/\Pg\bar{q}$-initiated subprocesses, which arise at $\order{\alphas\alpha^6}$, have a large and negative\footnote{The $\Pg q/\Pg\bar{q}$-contribution 
can be negative owing to the collinear counterterm necessary to cancel initial-state infrared divergences.} cross section in the third bin, 
and significantly contribute to the observed gap. 

The EW corrections are smaller, and the NLO EW distribution follows the LO curve more closely. 
Again, the corrections are negative at lower energies but fairly small in the first bins, 
as opposed to their QCD counterpart. Although the phase-space suppression of the real contribution 
also occurs in this case, its effect is smaller owing to the difference between the EW 
and QCD couplings. The EW corrections become positive at around $2300\GeV$, 
after which the NLO EW differential cross section departs from the LO one as the energy grows. 

The tails of the NLO QCD and EW distributions also behave differently. While the NLO EW curve shows a more uniformly falling tail, 
a change is visible in the rate at which the NLO QCD distribution diminishes after about $2\TeV$. This behaviour can be 
traced back to the real contributions of subprocesses of the $qq/\bar{q}\bar{q}$~type, which can bypass our cuts and become more important 
at larger energies, slowing down the decay of the NLO QCD distribution
(see also \refse{sec:nlo2}).
All in all, the total NLO distribution is dominated by the QCD correction starting at around $500\GeV$. 

The middle and bottom right panels of \reffi{fig:cms_bytype} display the relative NLO EW and QCD cross sections of each subprocess type $i$, defined as
\begin{align}\label{eq:delta_bytype}
        \delta^{(i)}_{\rm EW/QCD} = 
        \frac{\sigma_{\rm LO}^{(i)} + \delta\sigma_{\rm NLO \; EW/QCD}^{(i)}}{ \sigma_{\rm LO}^{\rm tot} + \delta\sigma_{\rm NLO \; EW/QCD}^{\rm tot} } \;.
\end{align}
Here, $\sigma_{\rm LO}^{(i)}$ and $\delta\sigma_{\rm NLO}^{(i)}$ stand for 
the LO cross section of the $i$-th subprocess type and the corresponding NLO correction, which contains only the higher-order contribution. 
The superscript ``tot'' refers to the sum over $i$.
%We remark that not all subprocesses contribute to the cross section at $\order{\alpha^6}$ (see Table \ref{tab:LO2}). 
%In such cases, $\sigma_{\rm LO}^{(i)} = 0$ in Eq.(\ref{eq:delta_bytype}). 

The middle panel on the right shows that the NLO EW correction is dominated by contributions of $q\bar{q}$ type for energies below $3\TeV$, 
while $\gamma q/\gamma\bar{q}$ contributions become larger for higher energies. We attribute this behaviour to the high-energy enhancement of 
the photon PDF. This contribution is responsible for the rise of the
NLO EW above the LO in the upper right plot. Relative contributions to other subprocess types do not exceed $5\%$ in the shown range.

On the bottom panel of the right plot in \reffi{fig:cms_bytype}, we observe an interesting behaviour in the third bin,  
between 300 and $450\GeV$. There, the relative NLO cross section $\delta^{(q\bar{q})}_{\rm QCD}$ is $364\%$, 
while $\delta^{(\Pg q)}_{\rm QCD}$ is $-274\%$ (not visible in the plot). 
These large values are ultimately due to the large and negative NLO QCD corrections, visible on the top panel, 
which lead to a relatively small denominator in Eq.~(\ref{eq:delta_bytype}) for that bin.
The rise of the QCD corrections for large
$\langle\sqrt{\hat{s}}\rangle$ is due to real gluon corrections that
evade the invariant-mass cut in \refeq{eq:setup2}.

\begin{figure}
        \centering
        \includegraphics[width=0.505\textwidth,page=1]{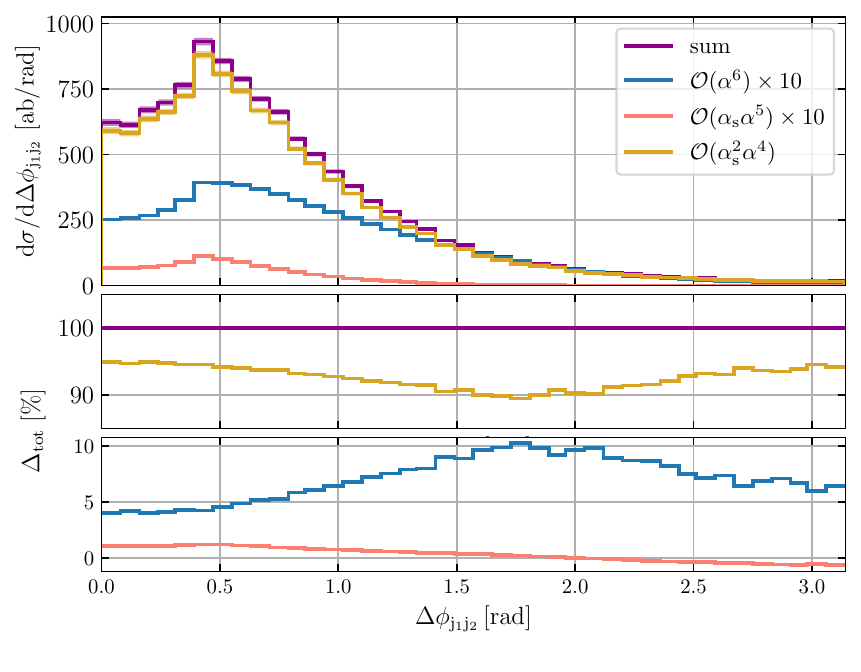}%
        \includegraphics[width=0.495\textwidth,page=1]{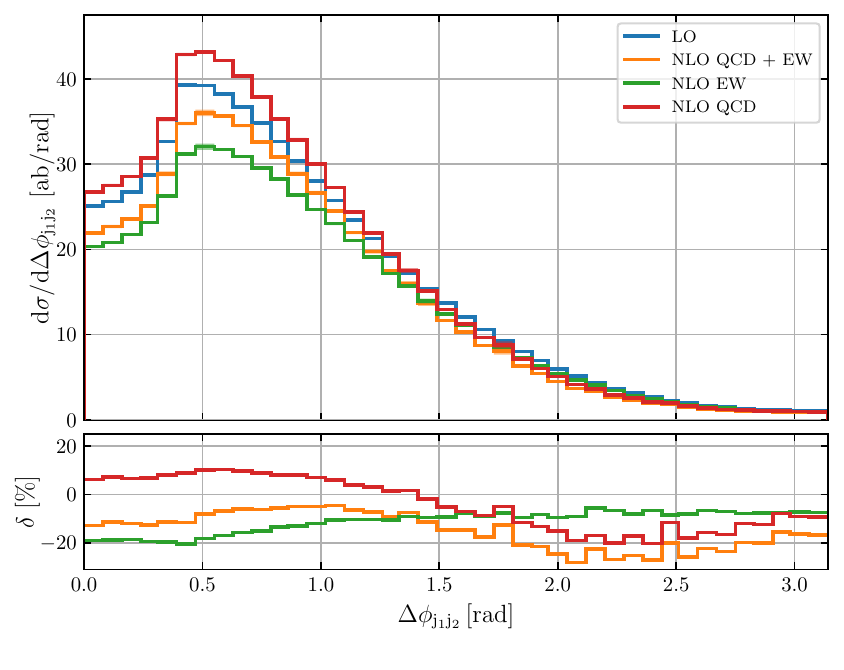}%
        \\
        \includegraphics[width=0.5\textwidth,page=8]{plots/paper/LO_byorder}%
        \includegraphics[width=0.5\textwidth,page=8]{plots/paper/NLO_byorder}%
        \caption{Distributions in the azimuthal angle between the tagged jets, $\Delta\phi_{\Pj_1 \Pj_2}$, (top row) 
                and their rapidity difference, $\Delta y_{\Pj_1 \Pj_2}$, (bottom row) at LO (left column) and NLO (right column). 
                %               The top panel of the left plot display the different leading orders, as well as their sum. 
                %               In plots where a linear scale for the vertical axis is employed, some curves are rescaled 
                %               as stated in the legends for visibility. 
                %               The lower panels of the left plots display the cross section of each leading order 
                %               relative to their sum in percent.
                %               On the right plot, the middle (bottom) panel displays the NLO EW (QCD) cross section of each type of 
                %               partonic subprocess relative to the total NLO EW (QCD) cross section.
              See the main text for a description of the individual plots.
        }
        \label{fig:byorder1}
\end{figure}

\begin{figure}
        \includegraphics[width=0.5025\textwidth,page=5]{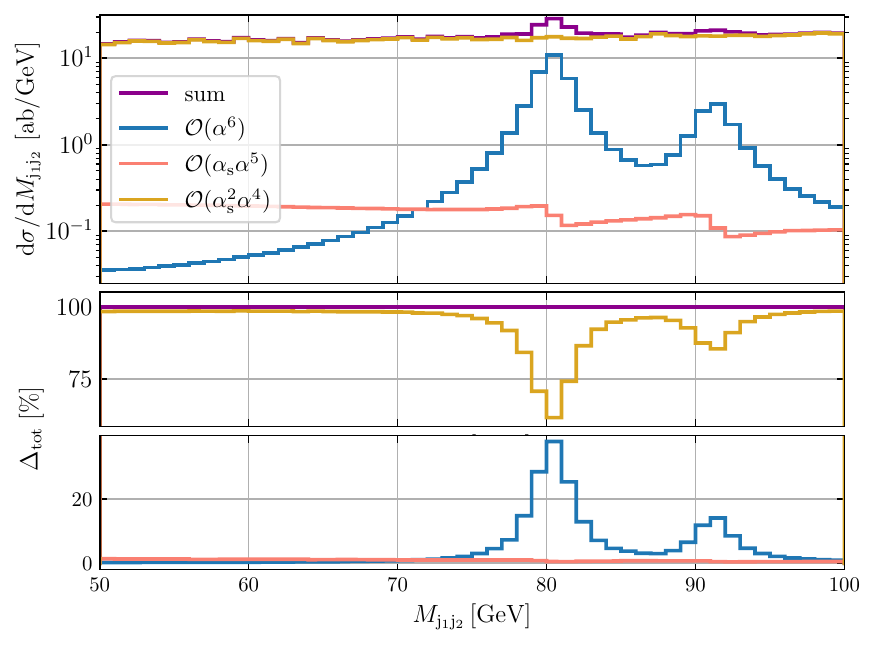}%
        \includegraphics[width=0.4975\textwidth,page=5]{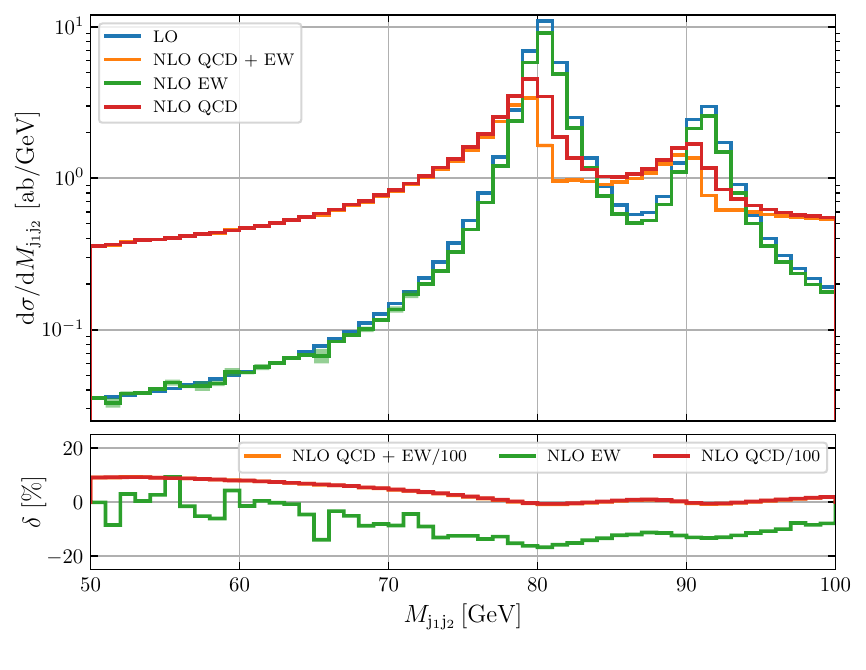}%
        \\
        \includegraphics[width=0.5025\textwidth,page=10]{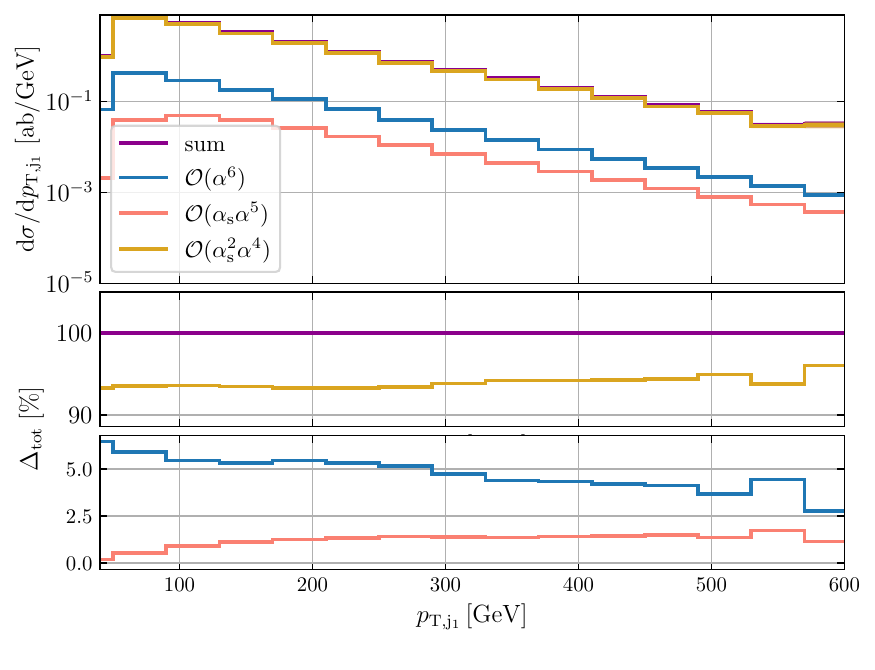}%
        \includegraphics[width=0.4975\textwidth,page=10]{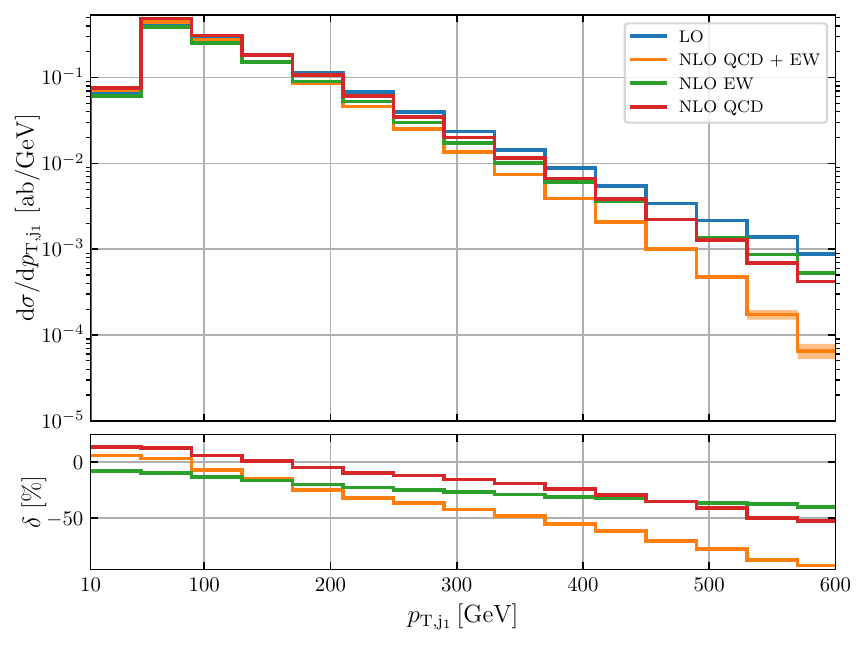}%
        \caption{Distributions in the invariant mass of the tagged-jet pair, $M_{\Pj_1\Pj_2}$, (top row) and the transverse momentum
                of the hardest tagged jet, $p_{\rm T, \Pj_1}$, (bottom row).
        }
        \label{fig:byorder2}
\end{figure}

\begin{figure}
        \includegraphics[width=0.503\textwidth,page=3]{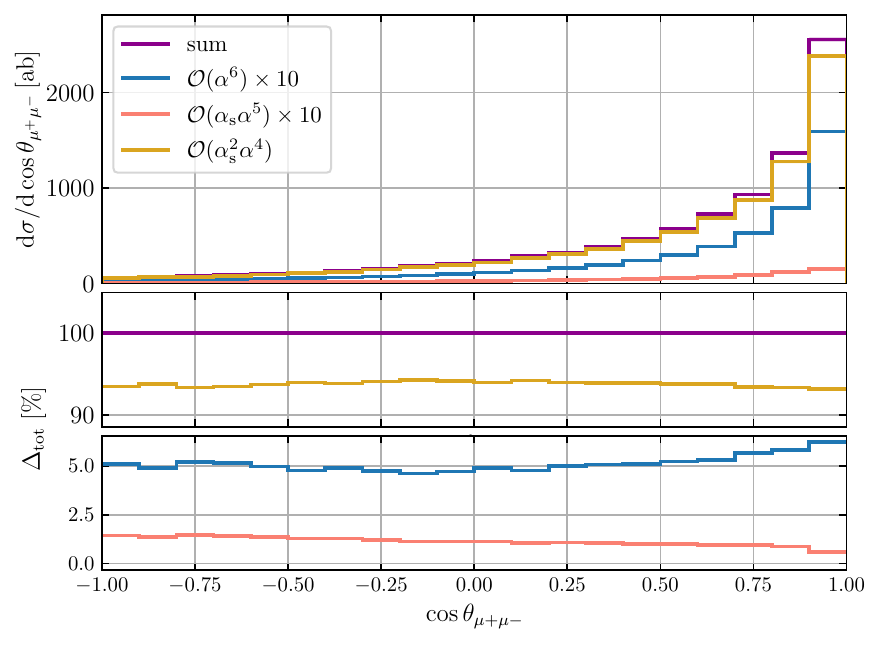}%
        \includegraphics[width=0.497\textwidth,page=3]{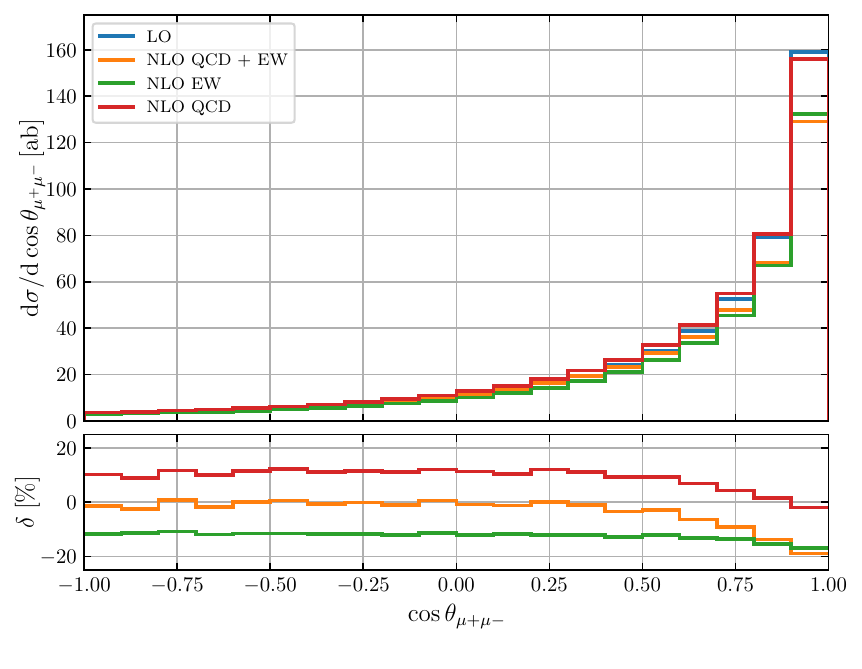}%
        \\
        \includegraphics[width=0.5\textwidth,page=9]{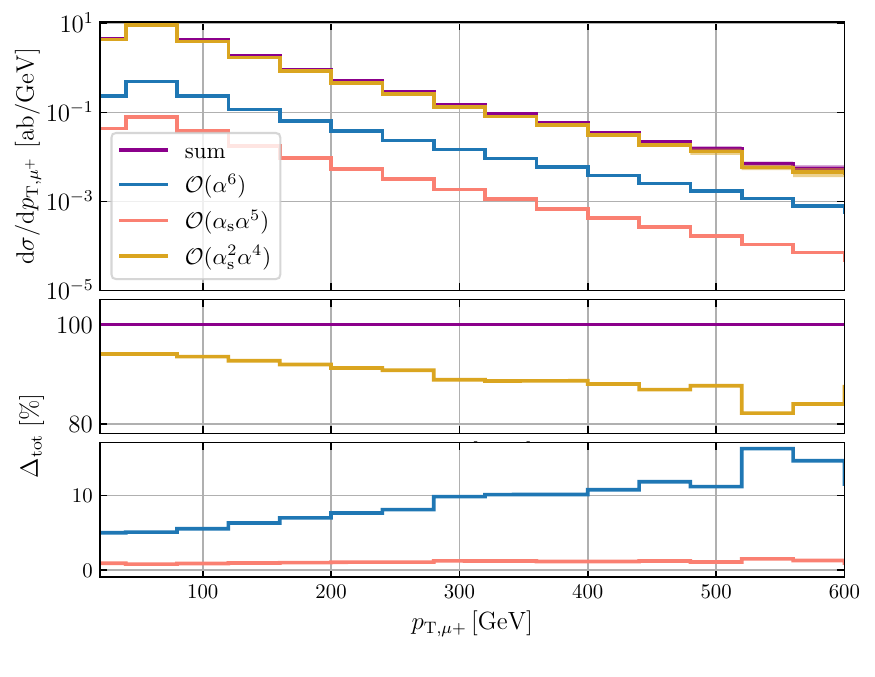}%
        \includegraphics[width=0.5\textwidth,page=9]{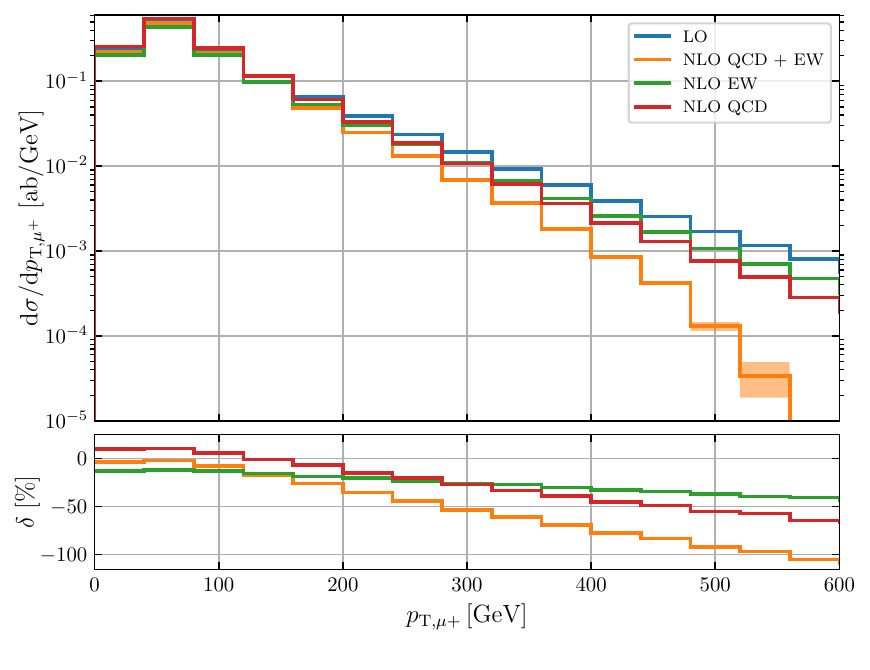}%
        \caption{Distributions in the cosine of the opening angle
          between the muon and antimuon, $\cos\theta_{\mu^+\mu^-}$, (top row)
                and the transverse momentum of the antimuon, $p_{\mu^+}$, (bottom row).
        }
        \label{fig:byorder3}
\end{figure}

\begin{figure}
        \includegraphics[width=0.5\textwidth,page=4]{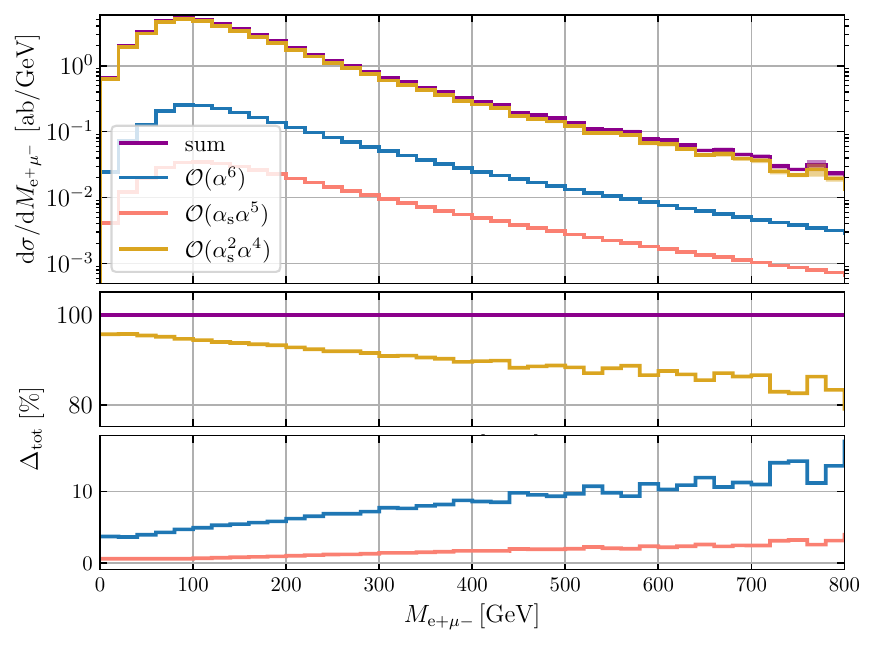}%
        \includegraphics[width=0.5\textwidth,page=4]{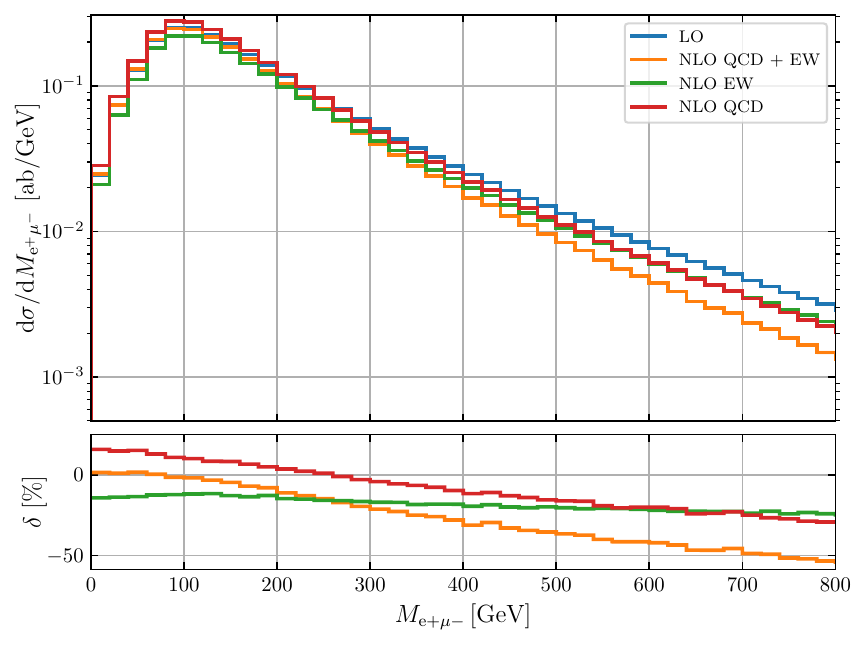}%
        \\
        \includegraphics[width=0.5\textwidth,page=7]{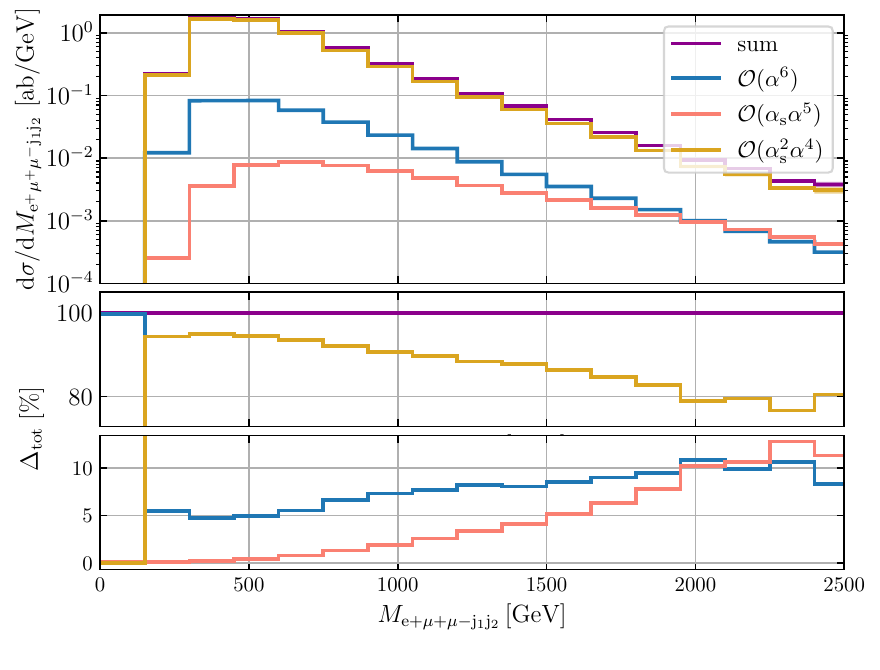}%
        \includegraphics[width=0.5\textwidth,page=7]{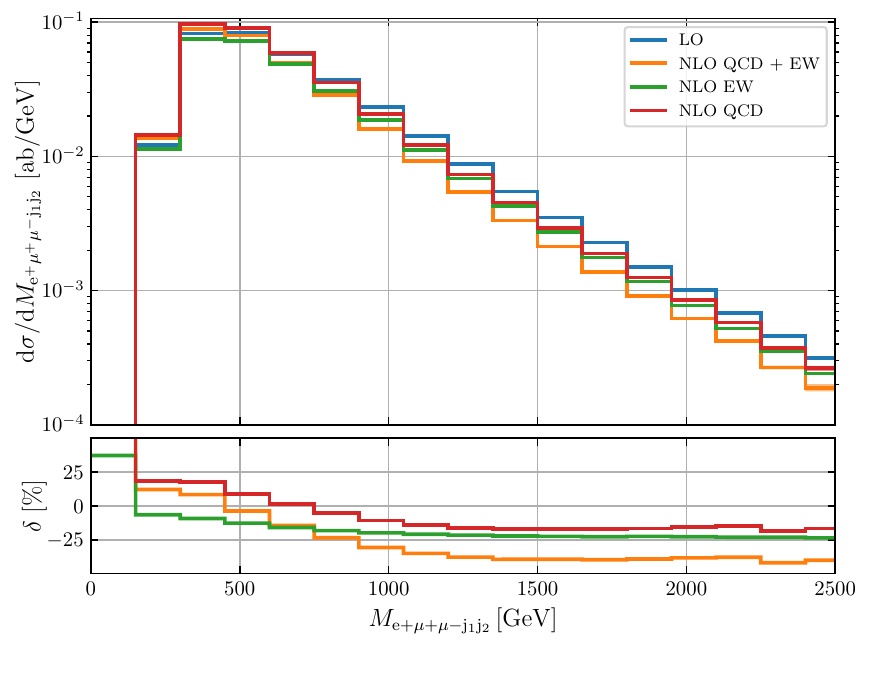}%
        \caption{Distribution in the invariant mass of the positron--muon system, $M_{\Pe^+\mu^-}$ (upper row), and the invariant mass
                of the system composed of all visible particles, $M_{\Pe^+\mu^+\mu^-\Pj_1\Pj_2}$, (bottom row). 
        }
        \label{fig:byorder4}
\end{figure}

We continue by presenting some observables that are relevant for tri-boson studies. 
Figures~\ref{fig:byorder1}--\ref{fig:byorder4} display a selection of differential distributions at the 
calculated leading and next-to-leading orders.
%Here, only the contributions of the subprocess types listed in Table~\ref{tab:NLO_perc} are taken into account.
%
In each case, the top panel of the left plot displays the distribution at each LO as well as their sum (dark magenta). Whenever a linear scale is used, 
the curves of $\order{\alpha^6}$ (blue) and $\order{\alphas\alpha^5}$ (salmon) are rescaled for visibility. If present, the rescaling factor is 
specified in the legend. The QCD background is depicted in yellow. In the lower panel, the left plots show the relative contributions of each order
in percent, zoomed into separate ranges for visibility. 
The right plots contain distributions for the LO signal (blue) as well
as the corresponding corrections at NLO EW (green), NLO QCD (red), and total NLO (orange).
The relative NLO corrections in the middle and bottom panels are defined as
\begin{equation}\label{eq:delta_byorder}
        \delta = \frac{\delta\sigma_{\rm NLO \; EW/QCD}}{\sigma_{\rm LO}} \;,
\end{equation}
where $\sigma_{\rm LO}$ includes only contributions of $\order{\alpha^6}$.

We begin by considering jet observables. 
The upper row of \reffi{fig:byorder1} shows the distribution in
$\Delta\phi_{\Pj_1 \Pj_2}$, the azimuthal angle between the tagged jets. 
These jets mostly originate from the decay of a vector boson. Therefore, in the laboratory frame we expect them to have a small angular separation, 
resulting in an enhancement of the $\Delta\phi_{\Pj_1 \Pj_2}$ distribution towards smaller values. 
Indeed, on the upper panel of the left plot, all distributions rise towards smaller angle differences but peak at around $0.4$. 
This is related to the $R$-parameter of our jet-clustering algorithm. At LO, the requirement of two tagging jets can only be fulfilled if the distance
\begin{equation}\label{eq:distance_taggingjets}
        \Delta R_{\Pj_1 \Pj_2} = \sqrt{ \Delta\phi_{\Pj_1 \Pj_2}^2 + \Delta y_{\Pj_1 \Pj_2}^2 } \,
\end{equation} 
between them in the $\phi$--$y$ plane is larger than 0.4, which effectively acts as a cut.
If $\Delta\phi_{\Pj_1 \Pj_2} < 0.4$, then Eq.~(\ref{eq:distance_taggingjets}) forces $\Delta y_{\Pj_1 \Pj_2}$ to assume larger values, 
reducing the available phase-space volume. 
As visible in the lower left panel, the relative contribution of the
$\order{\alpha^6}$ varies between $4\%$ and $10\%$ with a maximum near
$\Delta\phi_{\Pj_1 \Pj_2}=1.75$, while the relative interference
contribution of order $\order{\alphas\alpha^5}$ falls from $1.2\%$ to
$-0.6\%$ at $\Delta\phi_{\Pj_1 \Pj_2}=\pi$. Accordingly the dominant
$\order{\alphas^2\alpha^4}$ varies between $95\%$ and $89\%$. 

The shapes of the absolute NLO distributions of $\Delta \phi_{\Pj_1 \Pj_2}$ on the upper panel of the top right plot of \reffi{fig:byorder1} 
are similar to the LO distribution. 
On the bottom panel, we recognise a difference in behaviour between NLO EW and QCD corrections. 
While the former is consistently negative and decreases in size from
$-21\%$ to $-7\%$ towards higher values of $\Delta \phi_{\Pj_1 \Pj_2}$, 
the QCD correction amounts to $6\%$ at $\Delta\phi_{\Pj_1 \Pj_2}=0$,
grows to $10\%$ at 0.6, where the bulk of the cross section sits, and
then goes down to $-20\%$ for $\Delta\phi_{\Pj_1 \Pj_2}\gtrsim2$. 
%before going back to $-9.3\%$.
Since the NLO EW corrections are larger than the QCD ones in the bulk of the distribution, the total NLO correction  ranges between $-5\%$ and $-28\%$.

The lower row of \reffi{fig:byorder1} shows the distribution in the
rapidity difference between the tagged jets $\Delta y_{\Pj_1 \Pj_2}$. 
The distribution is symmetric with two peaks at $\pm0.4$ as in the
$\Delta \phi_{\Pj_1 \Pj_2}$ distribution. 
The relative contribution of the $\order{\alpha^6}$ amounts to
$5$--$6\%$ and the interference contribution reaches at most $1\%$.
The relative QCD and EW corrections, in the right plot, are rather flat where the
distribution is sizeable. 
For rapidity differences larger than $1.7$, the QCD correction becomes
large, reaching close to $300\%$ at $\Delta y_{\Pj_1 \Pj_2}=\pm2$ (not visible in the figure). 
This behaviour is not unexpected, since contributions compatible with VBS, for which larger rapidity differences are typical, 
can evade our tri-boson selection cuts at $\order{\alphas\alpha^6}$ via the real emission, as discussed in \refse{sec:nlo2}. 
The distributions in $\Delta y_{\Pj_1 \Pj_2}$ vanish for values larger
than $2.1$ as a consequence of the cuts in \refeqs{eq:setup1} and
\eqref{eq:setup2} on $M_{\Pj_1\Pj_2}$ and $p_{\text{T},\Pj_i}$, which
constrain the value of $\Delta y_{\Pj_1 \Pj_2}$ according to
Eq.~(49.45) of \citere{Workman:2022ynf} 
\begin{align}\label{eq:mjj}
        M^2_{\Pj_1\Pj_2} = 2 \, p_{T,\Pj_1} \, p_{T,\Pj_2} \left( \cosh \Delta y_{\Pj_1\Pj_2} - \cos\Delta \phi_{\Pj_1\Pj_2} \right) \; .
\end{align}

Figure \ref{fig:byorder2} shows in its upper row the distribution in the invariant mass of the tagged-jet pair $M_{\Pj_1\Pj_2}$. 
On the left plot, the LO signal clearly displays the expected peaks at
the $\PW$- and $\PZ$-boson masses, which are even visible in the sum
of leading orders. The QCD background slightly increases with growing
invariant mass, while the interference contribution tends to decrease
with two steps at the vector-boson masses. 
The lower panel shows that the LO signal represents $38\%$ of the total LO cross section at the $\PW$-boson peak and $14\%$ at the $\PZ$-boson peak. The interference
contribution remains below $1.5\%$ in the whole considered range.
On the right plot, the left flanks of both peaks receive large,
positive QCD corrections of up to $1000\%$ (note that the curves for the
real QCD correction is scaled down by a factor $100$). The peaks themselves are reduced and 
slightly shifted towards lower values in the NLO QCD distribution, and the first bins to the right of each peak centre have negative QCD corrections. 
This radiative-return effect is well known and due to the real gluon emission from a tagged jet. If the gluon is well separated, 
it carries away some of the energy from the products of the
hadronically decaying vector boson, resulting in a smaller invariant mass of the tagged jets. 
The right flank of the $\PZ$-boson peak receives substantial positive QCD corrections. These are also driven by the additional real radiation, 
which can influence the jet identification and thereby increase the likelihood for the invariant mass of the tagged jets to be further above the $\PZ$-boson resonance~\cite{Ballestrero:2018anz,Denner:2020zit,Denner:2024ufg}. 
Specifically, the extra QCD parton can be tagged as a jet, while the
other tagging jet results from the recombination of the weak-boson
decay products into a single jet, shifting the resonant contributions
to higher invariant masses. 
Alternatively, the extra real parton and one quark from the vector-boson decay may be recombined into a tagged jet, while the other quark from the decay is tagged as the other jet.

The behaviour of the NLO EW distribution is significantly different. It is consistently negative and takes its largest value of $-17\%$ at the $\PW$ peak. 
The radiative-return effect owing to unclustered real-photon radiation
is moderate for the distribution in $M_{\Pj_1\Pj_2}$. This is due
to the smallness of the EW coupling relative to the strong coupling
and the relative smallness of the quark charges with respect to the
lepton charges.

The distribution in the transverse momentum of the hardest jet, $p_{\rm T, \Pj_1}$, is shown on the lower row of \reffi{fig:byorder2}. 
At LO, the relative contribution of the LO signal decreases with
growing $p_{\rm T}$ from $6.4\%$ to $2.6\%$ at
$p_{\rT,\Pj_1}=600\GeV$, while the LO interference grows from $0.2\%$ to $1.8\%$. 
Although the relative total NLO correction (right plot) amounts to
only $6\%$  for small transverse momenta, it significantly influences
the tail of the distribution, reaching $-87\%$ at $600\GeV$.
The relative EW correction varies from $-8\%$ to $-37\%$ with increasing transverse momentum. 
This increase in size towards higher energies is driven by the large EW Sudakov logarithms mentioned in \refse{sec:integrated}. 
The QCD correction falls from positive values at
small transverse momentum down to $-52\%$ at $600\GeV$. 
We attribute these large negative QCD corrections at high transverse
momenta to our choice of a fixed scale at a low value [cf.~\refeq{eq:scaleA}]. 
The appearance of large negative EW and QCD corrections is common to all
distributions for high transverse momenta or invariant masses. 

We now examine distributions in observables related to the charged leptons. 
On the top row of \reffi{fig:byorder3}, we present the distribution in the cosine of the angle between the muon and the antimuon, $\cos\theta_{\mu^+\mu^-}$. 
Both the left and the right plots display similar behaviours for all curves, with a small opening angle being preferred in all cases. The relative LO differential cross sections in the
lower panel to the left do not vary by more than $2\%$. The relative NLO EW correction on the lower right panel is between $-10.5\%$ and $-12\%$ in the range from $-1$ to $0.5$ 
but increases in size towards $\theta_{\mu^+\mu^-} = 0$ reaching $-17\%$ in the rightmost bin. 
This results from the fact that at higher energies, where the
EW Sudakov corrections are large, the $\PZ$ bosons are more boosted,
leading to more collinear muon--antimuon pairs.
The QCD correction remains between $9\%$ and $12\%$ for
$\cos\theta_{\mu^+\mu^-}$ within  $-1$ and $0.6$. 
Towards larger values, it falls and becomes $-2\%$ in the last bin,
again because this region is dominated by contributions from high
scattering energies. 
Up to $\cos\theta_{\mu^+\mu^-} \approx 0.4$, the EW and QCD corrections largely cancel giving a total NLO correction 
between $-2.5\%$ and $0.7\%$. Thereafter, the total NLO correction is dominated by EW corrections and almost reaches $-19\%$ in the last bin.

The distribution in the transverse momentum of the antimuon is shown on the lower row of \reffi{fig:byorder3}.  
The relative $\order{\alpha^6}$ contribution grows towards larger energies from $5\%$ to $15\%$, while the one of $\order{\alphas^2\alpha^4}$ decreases from $94\%$ to $84\%$. 
The total relative NLO corrections behave similar to those of
the $p_{\rm T, \Pj_1}$ distribution. They range between $-2\%$ and
$-4\%$ at low transverse momenta but become large in the tail, and even reach
$-105\%$ for  $p_{\rm T, \mu^+}=600\GeV$. 
The EW correction grows in size from $-13\%$ to $-41\%$ with
increasing transverse momentum. The QCD correction is $+10\%$ at
small $p_{\rT,\mu^+}$ and then becomes negative with increasing $p_{\rm T, \mu^+}$ to reach $-64\%$ at $600\GeV$.

We turn to \reffi{fig:byorder4}, which shows the distribution in the invariant mass of the positron--antimuon system $M_{\Pe^+\mu^-}$ on its upper row.
On the lower panel of the left plot, the LO signal increases from $3.6\%$ at small invariant masses to $14\%$ at $800\GeV$, 
while the interference grows from $0.6\%$ to $3.1\%$ in the same invariant-mass range. 
On the right plot, the positive QCD correction and the negative EW
correction approximately cancel to give a total positive relative NLO
correction of at most $2\%$ in the range below $100\GeV$, 
where the peak of the distribution is located. For larger values, the total NLO correction becomes negative owing to the decrease of the QCD correction, 
which itself turns negative at around $250\GeV$. As a consequence, the tail of the distribution obtains a significant suppression and the total NLO correction 
amounts to $-53\%$ at $800\GeV$. The relative EW correction only
slightly varies from $-12\%$ to $-24\%$ in the tail.

We finish our discussion with the lower row of \reffi{fig:byorder4}, which displays the distribution in the invariant mass 
of all visible final-state particles, $M_{\Pe^+\mu^+\mu^-\Pj_1\Pj_2}$. 
As in other LO results, the relative QCD background decreases towards
larger mass values, and in this case falls to $76\%$ at
$M_{\Pe^+\mu^+\mu^-\Pj_1\Pj_2}=2.5\TeV$. Correspondingly, the LO
signal increases from $5\%$ to $8\%$ and the interference from $0\%$
to $13\%$.
This is the only observable that we found, where the $\order{\alphas\alpha^5}$
contribution exceeds the LO signal in parts of phase space (above $2\TeV$). 
At NLO, both types of corrections follow a similar trend. For lower values of the invariant mass, both the QCD and the EW correction are positive. 
They both diminish for larger invariant masses, and end up being both negative and similar in size towards the upper limit of the shown range. In the tail of the distribution, the total NLO correction reaches $-42\%$.

%list of distributions
%Transverse momentum:
%\begin{itemize}
%       \item $\pt{\Pe^+},\; \pt{\mu^-},\; \pt{\mu^+},\; \pt{\nu_e}$
%       \item $\pt{j_1},\,\pt{j_2}$
%       \item $\pt{j_1j_2},\, \pt{\mu^+\mu^-}\,\pt{W^+}$
%       
%       \item $H_{\rm T\, \rm lep} = \sum_{i=1}^3\,\pt{\ell_i}$
%       \item $H_{\rm T\, \rm hadr}= \sum_{i=1}^2\,\pt{j_i}$
%       \item $H_{\rm T} = H_{\rm T\, \rm hadr}+H_{\rm T\, \rm lep}$
%\end{itemize}
%
%\noindent
%Rapidities and angles: $\left( R^2 = (\Delta y)^2 + (\Delta \phi)^2 \right)$
%\begin{itemize}
%       \item $y_{\Pe^+},\; y_{\mu^-},\; y_{\mu^+}$
%       \item $y_{j_1},\; y_{j_2}$
%       \item $\Delta y_{\mu^+\mu^-},\; \Delta y_{\Pe^+\mu^-},\; \Delta y_{\Pe^+\mu^+}$
%       \item $\Delta R_{\mu^+\mu^-},\; \Delta R_{\Pe^+\mu^-},\; \Delta R_{\Pe^+\mu^+}$
%       \item $\Delta y_{j_1j_2},\; \Delta R_{j_1j_2},\; \Delta\phi_{j_1j_2}$
%       \item $\cos{\theta_{\mu^+\mu^-}}$
%\end{itemize}
%
%\noindent
%Invariant and transverse masses:
%\begin{itemize}
%       \item $m_{\mu^+\mu^-},\,m_{\Pe^+\mu^-},\,m_{\Pe^+\mu^+}$
%       \item $m_{\Pe^+\mu^+\mu^-}$
%       
%       \item $m_{j_1j_2}$
%       
%       \item $m_{\PW^+\PZ}, m_{\PW^+\PZ j_1j_2}$
%       \item $\tm{\PW^+}$
%\end{itemize}

%%% Local Variables: 
%%% mode: latex
%%% TeX-master: "wvz_paper"
%%% End: 

%% file: conclusions.tex
\section{Conclusions}\label{sec:conclusions}

In this work we presented a calculation for the process
$\Pp\Pp\to\mu^+\mu^-\Pe^+\nu_\Pe\,\Pj\,\Pj$ at the LHC at $13.6\,$TeV in a phase space
constructed to enhance the tri--boson-production mechanism. The same process
was studied in~\citere{Denner:2019tmn} in a VBS-like phase space.

We computed the complete set of LO contributions. They comprise the $\mathcal{O}(\alpha^6)$,
which is the only one that includes the tri-boson signal, and the
$\mathcal{O}(\as\alpha^5)$ and $\mathcal{O}(\as^2\alpha^4)$, which are part of the background.
Additionally, we evaluated NLO corrections at $\mathcal{O}(\alpha^7)$ (NLO EW)
and $\mathcal{O}(\as\alpha^6)$ (NLO QCD). We point out that the latter contribution, which we refer to
as the QCD correction to our LO signal, also receives EW corrections to the $\mathcal{O}(\as\alpha^5)$.
All resonant and non-resonant effects, together with interference terms,
have been exactly retained. All partonic channels relevant for the description of the tri-boson
signal have been taken into account, including the photon- and $\bar{\Pb}\Pb$-induced ones. In our
calculation we only excluded partonic channels with bottom quarks in
the final state by assuming a perfect $\Pb$-jet tagging and
veto. These channels can
induce a top-quark resonance and highly contaminate our signal, as was confirmed by evaluating
them at LO. We found that they amount to roughly $25\%$ and $15\%$ of the
$\mathcal{O}(\alpha^6)$ at $\mathcal{O}(\as\alpha^5)$ and $\mathcal{O}(\as^2\alpha^4)$, respectively.

We presented results for the fiducial cross section and differential distributions. At the inclusive level we
performed a thorough study of the individual contributions of the different partonic channels
at the different perturbative orders.  When considering corrections to
the $\mathcal{O}(\alpha^6)$, the LO signal, NLO EW corrections amount to $-14\%$, and receive
contributions predominantly from quark--antiquark- and photon--quark-induced channels. These corrections
are considerably larger than those found for other tri-boson processes in previous
calculations~\cite{Nhung:2013jta,Dittmaier:2017bnh,Schonherr:2018jva,Dittmaier:2019twg,Denner:2024ufg}, and as
sizeable as the NLO EW corrections to the process in a VBS-prone
acceptance region~\cite{Denner:2019tmn}. This is in contrast to the outcome of
\citere{Denner:2024ufg},
where EW corrections to $\Pp\Pp\to\mu^+\nu_\mu\Pe^+\nu_\Pe\,\Pj\,\Pj$ were found to be much smaller
in a tri-boson-prone signal region than in a VBS-prone one. We attribute this to the comparably
high average partonic centre-of-mass energy of our process, which is
partly due to the chosen setup, which enhances the size of the
EW corrections through 
large EW logarithms of Sudakov type. On the other hand, NLO QCD corrections are moderate at the integrated level
and amount to $4\%$ of the LO signal.

At the differential level both NLO EW and QCD corrections can become sizeable and show an interesting interplay
for most of the observables that we studied. Not only does their inclusion affect the normalisation
of the predictions of the LO signal, but it also leads to relevant shape effects. For the majority
of the phase-space regions where the bulk of the cross section resides, the two corrections have opposite
signs. Specifically, NLO EW corrections are negative and for angular observables, like the
azimuthal angle or rapidity difference of the tag jets, larger in absolute value than the corresponding
QCD ones. In the considered transverse-momentum and invariant-mass distributions,
we observed large cancellations between the two corrections in the most-populated phase-space regions. For this second
class of observables we saw that NLO EW corrections stay negative when moving to high-energy values,
with a growth in size driven by large EW Sudakov logarithms and can
reach $-40\%$ in the considered phase-space regions.
On the other hand, the NLO QCD
contributions change sign towards the tails of these distributions, providing an additional source of
large and negative corrections which can amount up to $-60\%$. This latter behaviour is
understood as an artifact of the choice the \PW-boson mass as a fixed renormalisation
scale that is small compared to the energy scales in the distribution tails.

With this work we presented new results for $\Pp\Pp\to\mu^+\mu^-\Pe^+\nu_\Pe\,\Pj\,\Pj$
in a phase-space region that by now has received little attention from the theory community,
and we provided one more study of the triple-vector--boson-production
mechanism for a completely new final state.  The importance of this process in consolidating
our knowlegde of the SM and in constraining new-physics scenarios is established.
The improved theoretical control on this process will potentially be a crucial ingredient
  for the experimental analyses in the upcoming High-Luminosity stage of the LHC.

%%% Local Variables: 
%%% mode: latex
%%% TeX-master: "wvz_paper"
%%% End: 